\documentclass[
 aps,prl
 amsmath,
 amssymb,
 reprint,
 superscriptaddress,
 floatfix,
]{revtex4-2}

\usepackage{graphicx}
\usepackage{dcolumn}
\usepackage{bm}
\usepackage{mathrsfs}
\usepackage{xfrac}
\usepackage{hyperref}
\usepackage[mathlines]{lineno}
\usepackage{color}
\usepackage[normalem]{ulem}
\usepackage{amsmath}
\usepackage[usenames,dvipsnames]{xcolor}
\usepackage{braket}
\usepackage{comment}

\usepackage{float}

\begin{document}

\title[The Superconducting Quasiparticle-Amplifying Transmon: A Qubit-Based Sensor for meV Scale Phonons and Single THz Photons]{The Superconducting Quasiparticle-Amplifying Transmon: A Qubit-Based Sensor for meV Scale Phonons and Single THz Photons}

\author{C.W.~Fink}
\email[]{cwfink@lanl.gov}
\affiliation{Materials Physics and Applications -- Quantum, Los Alamos National Laboratory, Los Alamos, New Mexico 87545, USA}

\author{C.P.~Salemi}
\email[]{salemi@stanford.edu}
\affiliation{SLAC National Accelerator Laboratory, Menlo Park, California 94025, USA}
\affiliation{Kavli Institute for Particle Astrophysics and Cosmology, Stanford University, Stanford, CA 94035, USA}

\author{B.A. Young}
\affiliation{Santa Clara University, Santa Clara, CA 95053, USA}

\author{D.I. Schuster}
\affiliation{Department of Applied Physics, Stanford University, Stanford, CA 94035, USA}

\author{N.A. Kurinsky}
\email[]{kurinsky@slac.stanford.edu}
\affiliation{SLAC National Accelerator Laboratory, Menlo Park, California 94025, USA}
\affiliation{Kavli Institute for Particle Astrophysics and Cosmology, Stanford University, Stanford, CA 94035, USA}



\date{\today}

\begin{abstract}
With great interest from the quantum computing community, an immense amount of R\&D effort has been invested into improving superconducting qubits. The technologies developed for the design and fabrication of these qubits can be directly applied to applications for ultra-low threshold particle detectors, e.g. low-mass dark matter and far-IR photon sensing. We propose a novel energy resolving sensor based on the transmon qubit architecture combined with a signal-enhancing superconducting quasiparticle amplification stage. We refer to these sensors as SQUATs: Superconducting Quasiparticle-Amplifying Transmons. We detail the operating principle and design of this new sensor and predict that with minimal R\&D effort, solid-state based detectors patterned with these sensors can achieve sensitivity to single THz photons, and sensitivity to $1\,\mathrm{meV}$ phonons in the detector absorber substrate on the $\mu\mathrm{s}$ timescale.
\end{abstract}


\maketitle


\textit{Introduction---}The increasing maturity of superconducting qubits over the past few decades has allowed the field of superconducting quantum computing to flourish, producing a massive industry centered on the goal of improving quantum coherence. Breakthrough studies in the last few years have shown that environmental radioactivity can induce correlated errors in qubit arrays~\cite{mit_2020, Wilen_2021, McEwen_2021}, and subsequent work has demonstrated that charge noise and phonon-induced drops in coherence time can be mitigated by designing phonon sinks and reducing photon coupling to films nearest to the qubits~\cite{Martinis_2021, Iaia_2022}. 

In parallel, advances in detector technology that have built on this wave of qubit fabrication expertise have shown that this same technology can be applied to energy sensing at the THz (meV) scale. The first demonstration of the Quantum Capacitance Detector (QCD)~\cite{Shaw_QCD, echternach_2018} showed that single THz photon detection can be achieved by utilizing a Cooper pair box coupled to a resonator -- a structure analogous to many early charge-sensitive qubits. This demonstration and work to understand the radiation sensitivity of qubits have opened up a new regime of sensing leveraging the single quasiparticle (QP) sensitivity of qubit-derived structures. The main sensing mechanism comes from the quantized nature of the qubit and the charge sensitivity of the transition, such that a single QP tunneling across the junction is easily measurable in real time~\cite{Rist__2013, PhysRevLett.121.157701,Mannila_2021, Kurter_2022}. If energy can be focused into breaking Cooper pairs near the junction, as is done in Ref~\cite{echternach_2018}, then very low thresholds are achievable. 

In this letter we propose a novel sensing technology based on weakly charge-coupled transmon qubits, which we call SQUATs: Superconducting Quasiparticle-Amplifying Transmons. The scheme described in this paper differs from the QCD design~\cite{Shaw_QCD, echternach_2018} and that of conventional superconducting qubits~\cite{koch, PhysRevLett.111.080502,walraff_scqubit2004, Kelly_2015, PhysRevApplied.11.054072, Wilen_2021} by removing the readout resonator entirely from the architecture and relying on direct readout of the qubit transition frequency to detect tunneling events. By removing the readout resonator, we are no longer sensitive to the quantum capacitance as in the QCD, but rather we are sensitive to the quantum inductance in the non-linear LC resonator that makes up the qubit. This architecture change allows for significant reduction in the overall size of the unit cell, increases in pixel density, reduction of two-level system noise, and increase in detection efficiency for uniformly distributed radiation signals. We will show that these devices should outperform the energy sensitivity of competing technologies, while the RF based readout scheme allows them to be naturally multiplexed, thus allowing for highly pixelizable, ultra-low threshold single THz photon and single-phonon detection.


\textit{Operating Principle---}Transmon qubits~\cite{koch} are anharmonic LC oscillators where the capacitance comes from the proximity of metal islands that are connected by a Josephson junction (JJ), which provides a nonlinear inductance.  The nonlinearity of the inductance creates unequally spaced energy levels; typical qubit operation uses only the first two levels, the ground, $\ket{0}$, and first excited, $\ket{1}$, states.

The energy difference between these states, $E_{01}$, is dependent on the qubit's design, parametrized by the qubit's single-electron charging energy, $E_C=e^2/2C$, and the Josephson energy, $E_J=\hbar I_c/2e$.  Here $e$ is the electron charge, $C$ is the qubit capacitance, and $I_c$ is the critical current. In the transmon limit, which corresponds to $\xi \equiv \frac{E_J}{E_C} \gg 1$, we can model the qubit energy accurately as \cite{koch}
\begin{equation}
    E_{01} \approx \hbar\omega_{0} + \hbar\chi_0\cos(\pi n_g) \,,
    \label{eq:energy}
\end{equation}
where $n_g$ is the island charge and
\begin{equation}\label{eq:f0}
    \hbar\omega_0 \approx \sqrt{8E_CE_J}-E_C \,.
\end{equation}
 The magnitude of the charge dependence of $E_{01}$ is controlled by
\begin{equation}
    \frac{2\chi_0}{\omega_0} \approx e^{-\sqrt{8\xi}}\left[60.7\xi^{3/4}+5.37\xi^{1/4}\right] \,.\label{eq:dispersion}
\end{equation}

We can thus tune the qubit to a given frequency by careful design of the junction area and critical current density, as well as the total capacitance within the qubit geometry.

One feature of the nonlinearity of the relationship between $E_{01}$ and $E_C$ and $E_J$ is that the qubit energy has a periodic dependence on the effective number of charges on the island, $n_g$. Incrementing this charge by one electron switches the state parity as the argument of the cosine in Eq.~\ref{eq:energy} changes by $\pi$.  Note that $n_g$ includes both the continuous offset charge from the charge environment around the qubit as well as the discrete electrons that come from broken Cooper pairs.

\begin{figure}[H]
    \centering
\includegraphics[width=.95\linewidth]{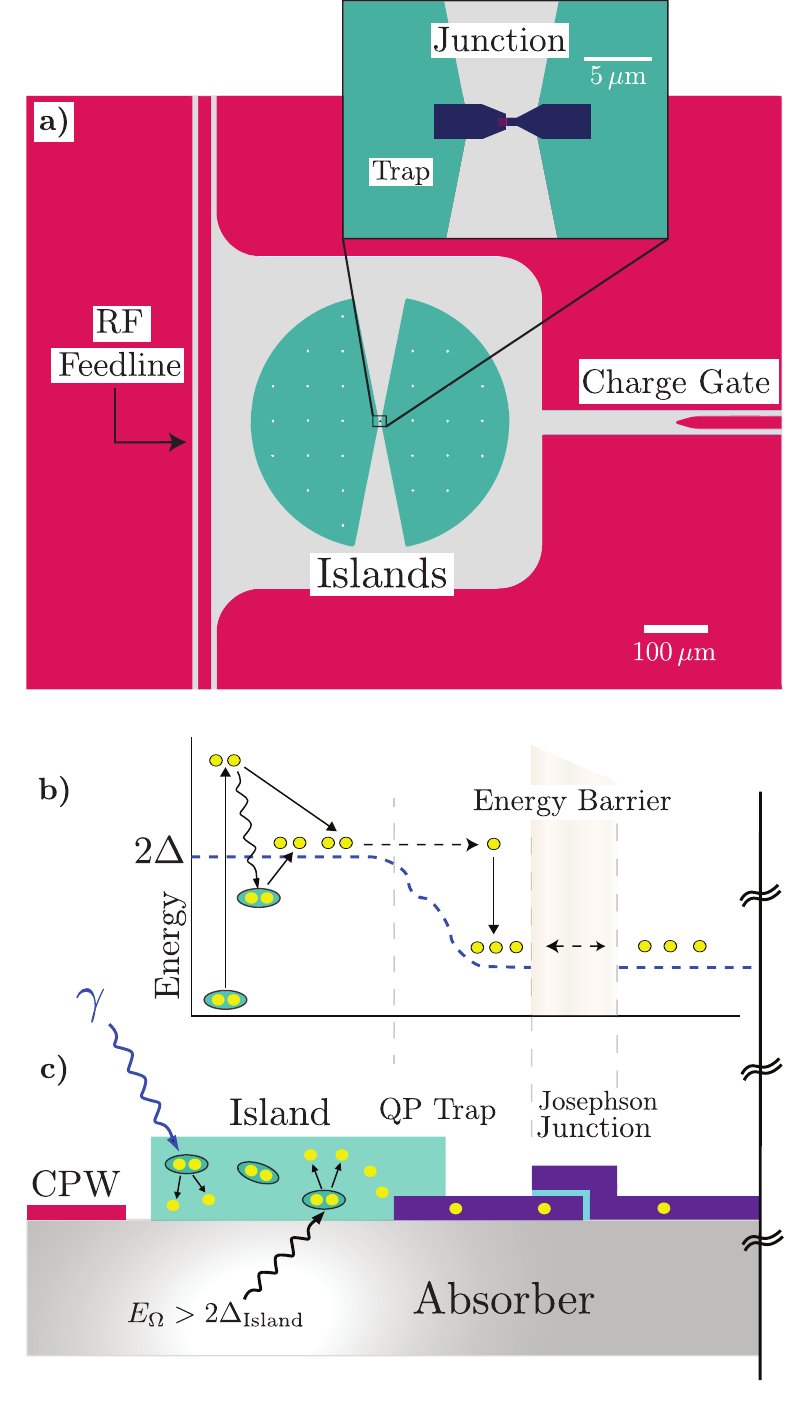}
    \caption{\textbf{a)} Photolithographic mask design for a first prototype SQUAT. The islands are composed of two sectors of a circle with island length ($\ell_\text{island}$) defined as the radius of the circle. The trap is the region of overlap between the island and the junction material. The length of the overlap is designed to be the same as the width such that the area of the trap can be parameterized with a single length parameter ($\ell_\text{trap}$). \textbf{b)} QP energy diagram showing signal measurement process. From left to right: A Cooper pair is broken by an incident particle with energy greater than $2\Delta_\text{Island}$ and creates QPs in an excited state, these QPs downconvert releasing phonons and lower energy QPs, these QPs diffuse until becoming trapped in the lower gap trap, the QPs then tunnel across the junction until they recombine or are trapped by an impurity. \textbf{c)} Cross-section of the sensor shown on top of a substrate (not to scale). Two possible signal paths are shown: direct absorption of a photon into the island (blue) or an athermal phonon from the substrate (black).}
    \label{fig:sensor_diagram}
\end{figure}

There are two ways to create non-equilibrium QPs in the qubit islands, as shown in Fig.~\ref{fig:sensor_diagram}. First, direct energy deposition in the island produces a pair of QPs in an excited state, which quickly generate a larger QP population as they settle to the superconducting energy gap. This is the case for photon absorption or direct collision of particles with the qubit island. Alternatively, athermal substrate phonons with energies large compared to the superconducting gap of the islands will similarly excite non-equilibrium QPs in the islands. Both processes lead to the same optimization for collecting the resulting QPs, but the two energy absorption methods require different considerations for external quantum efficiency, as described in the following sections. 

\textit{Phonon coupling---}Energetic electrons/holes or optical phonons created from an inital particle interaction in the substrate will rapidly downconvert to high energy acoustic athermal phonons. These phonons then undergo anharmonic decay until their mean free path is on the order of the characteristic size of the substrate~\cite{PhysRevApplied.11.064025}. The phonons will then travel ballistically in the substrate until being absorbed by the active (sensor) and passive (e.g. RF feedline, ground plane) metal films on the surface of the substrate. From~\cite{Fink_thesis}, the characteristic time scale for these phonon event signals can be approximated by 
\begin{align}
    \tau_\text{phonon} \approx \frac{4V_\text{det}}{\sum \left<c_\text{det}\right>f_{abs}^i A_i}
    \label{eq:tau_ph}
\end{align}
where $V_\text{det}$ is the volume of the substrate, $\left<c_\text{det}\right>$ is the average sound speed in the substrate, $A_i$ is the area of the $i^{th}$ absorbing material on the detector surface, and $f_{abs}^i$ is the phonon transmission probability between the substrate and the absorbing material. This transmission probability is modeled simply as
\begin{equation}
    f_{abs}^i = 1-\exp\left[\frac{2 h^i}{c_s^i \tau_B^i}\right], 
    \label{eq:abs}
\end{equation}
where $h^i$ is the absorber film thickness, $c_s^i$ is the sound speed in the film, and $1/\tau_B^i$ is the QP pair breaking rate, all for the $i^{th}$ film. In our design, the RF feedline and ground plane will be made of Nb, and the majority of the sensing material will be Al, values of $1/\tau_B^i$ for both of these materials can be found in~\cite{Kaplan_qp}.

There are two signal efficiency penalties that must be accounted for in this step. The first is, for a given interaction from an incident particle with an energy greater than the ballistic energy cutoff, the percentage of ballistic phonons with energy greater than twice the superconducting bandgap ($\Delta$) of the sensor material, which is typically $> 98\%$ for common materials~\cite{pyle_thesis}, e.g Si substrate instrumented with Al sensors. For this reason, we do not consider this efficiency at this point in our simulations. Secondly, we must account for the percentage of phonons that are absorbed by the non-instrumented areas of the detector surface. This efficiency factor can be calculated as the sum of the active (sensor) phonon absorption probability and area, divided by the total phonon absorption probability and area of every phonon absorbing surface:

\begin{equation}
    \varepsilon_\text{abs} = \frac{\sum^\text{active} f_{abs}^i A_i}{\sum^\text{total} f_{abs}^i A_i},
    \label{eq:phonon_eff}
\end{equation}

From the unit cell for our sensor shown in Fig.~\ref{fig:sensor_diagram}, we expect this can be placed in an array in such a way that $\sim5\%$ of the surface area of the entire chip will be active instrumented sensor area. The (passive) coplanar waveguide (CPW) feedlines also will cover an additional $\sim5\%$ of the detector surface. Lastly, a (passive) parquet patterned Nb grid ground plane similar to that used in~\cite{HVeV} will cover another $\sim5\%$ of the surface.  Despite the significantly larger superconducting critical temperature ($T_c$) of Nb vs. Al, preliminary studies suggest that a large fraction of the athermal phonon population in Si will be greater than twice the Nb superconducting gap and thus the Nb ground plane can become a significant source of phonon loss~\cite{ZoeLTD20}.

The remaining $\sim 85\%$ of the surface area will be left bare. Various surface treatments (e.g. polishing, chemical etching, etc) will be used in the fabrication R\&D to minimize phonon losses at the surfaces. Using these fill factors for the passive and active areas, and weighting by the phonon transmission probabilities in Eq.~\ref{eq:abs}, we thus find from Eq.~\ref{eq:phonon_eff} that we expect a phonon collection efficiency of $\varepsilon_\text{abs}\approx 37\%$. From Eq.~\ref{eq:tau_ph} the phonon collection time depends linearly on the detector thickness, and for a $500\,\mu\mathrm{m}$ thick Si wafer substrate, we expect $\tau_\text{phonon}\approx 2\,\mu\mathrm{s}$. If the signal loss from passive surfaces proves to be a problem, in a future iteration of this design we propose to use a flip-chip process in which the CPW qubit readout and control circuitry is on a separate substrate, similar to what is done in~\cite{Rosenberg_2017}. By doing this, nearly 100\% of the passive material can be removed from the detector chip.


\textit{Photon coupling---} In addition to athermal phonon measurement, these sensors can be used as single photon sensors via direct absorption in the islands or coupling of antenna mode to impedance-matched junctions (see, e.g.,~\cite{Houzet2019, Liu2024}). The shape of the sensor (see Fig.~\ref{fig:sensor_diagram}) is designed to maximally collect the induced QP signal, as described in the following text; additionally, this shape naturally lends itself to being engineered into a broadband bow-tie antenna. Analogous to the phonon collection case, $\varepsilon_\text{abs}$ is now defined as the internal quantum efficiency for photon absorption, and the characteristic signal is a delta function impulse of energy into the sensor, rather than the exponential signal with a time constant defined in Eq.~\ref{eq:tau_ph}.

While a detailed simulation of the antenna structure is beyond the scope of this paper and left for future work, we note that similarly designed antenna micro-structures have achieved broadband absorption in the 1-10\,THz range~\cite{Ren2008, Apriono2018, Wahyudi2017}. For example, \cite{Apriono2018} has shown with simulations $10-100\,\mu\mathrm{m}$-size on-chip bowtie antennas with 20\% bandwidth around 1\,THz with 90\% efficiency and \cite{Wahyudi2017} has demonstrated a single-structure bowtie antenna that functions over 1-10\,THz.  Various studies of THz absorption in general thin films have also been done, with wide-ranging efficiencies \cite{Carelli2017,Mori2014,Grbovic2011,Naftaly2011}.  QCDs have achieved an optical efficiency of 90\% in a narrow band around 1.5\,THz using a mesh antenna absorber \cite{echternach_2018}.

The SQUAT photon absorption efficiency can also be improved by using thicker films, optimizing fabrication methods and material choice, applying surface treatments, and adding a reflector under the substrate to give photons a second pass through the device.  From the literature, we conservatively estimate that we can achieve a photon absorption efficiency of $\varepsilon_\text{abs}\approx 50\%$ photon absorption in Al over a 20\% bandwidth at THz frequencies \cite{Apriono2018,Stutzman2012,Wahyudi2017}.

\textit{Quasiparticle Collection Efficiency---} To maximize the sensitivity of the SQUAT, we must optimize its efficiency to collect QPs.  For the SQUAT signal path, `collecting a QP' means that it tunnels across the junction. A novel feature of the SQUAT design is that it acts as a QP multiplier by increasing the number of QPs through downconversion in the trap and through each QP having multiple tunnels, which is further increased by being trapped near the junction.  These compounding effects greatly enhance the signal. 

The geometry of the coplanar capacitor islands in our device is designed to both maximize surface fill factor and QP collection efficiency. This is optimized in three ways. 1) The geometry of the islands is set by the QP diffusion length of the material such that QPs have a long enough lifetime to reach the junctions. 2) We add QP traps around the junction to provide additional collection efficiency.  QP trapping~\cite{booth_qp_trap, PhysRevLett.64.954, ULLOM199698}, similar to that done with Quasiparticle-trap-assisted Electrothermal-feedback Transition-edge sensors (QETs)~\cite{qet}, can be achieved by fabricating the islands with a material with a larger $T_c$ than the junction. 3) With a sufficient difference in the $T_c$'s of the island and junction material, the trapping effect also results in a QP multiplication stage, as originally proposed by Booth for superconducting tunnel junctions (STJs)~\cite{booth_qp_trap}. 

For this proposed device, we choose Al for the athermal phonon collection region.  For the trapping region and junctions, we target a material with a superconducting gap that is roughly a factor of 10 lower than Al. There are a variety of options for materials that have $T_c$'s in the desired range of $\sim 100\,\mathrm{mK}$ and have been shown to grow nice oxide layers for junctions, e.g Hf~\cite{Hf_tes, KIM2012667}, Ti~\cite{Fukuda:11}, and AlMn~\cite{Deiker_2004, AlMn_nist, AlMn_NIS, 2016JLTP..184...66L}. For the remainder of this concept paper, we will refer to this lower gap material as `LG'. The simplest material to use would be AlMn (which we will assume for our calculations here), since its $T_c$ in thin films is easily tunable by changing the Mn content and it lends itself well to Josephson junction fabrication, as it naturally creates an aluminum oxide layer.

The full QP diffusion model is described in detail in Appendix~\ref{appendix:qp_model}, and we highlight the key features here. For a schematic diagram of the following, see Fig.~\ref{fig:sensor_diagram}.
\begin{enumerate}
    \item \emph{Phonon collection and initial QP downconversion in island--} When a phonon or photon of energy greater than $2\Delta_\text{island}$ enters the island, a Cooper pair will be broken with a probability given by Eq.~\ref{eq:abs}, and the resulting QPs will be promoted to well above the bandgap energy. These QPs will undergo a downconversion process to lower energy QPs and phonons, eventually yielding a semi-stable population of QPs at the band edge after $\mathcal{O}(1)\,\mathrm{ns}$~\cite{PhysRevB.61.11807}. During the downconversion process, a portion of the initial event energy will be lost as sub-gap phonons leak into the substrate~\cite{Kaplan_qp}. For conventional S-wave superconducting films, incident events with energy $E_\Omega \gtrsim 4\,\Delta_\text{island}$ will have approximately $60\,\%$ of the event energy remain in the QP system~\cite{Guruswamy_2014}. We will assume that the initial primary phonon/photon events will be in this energy regime and will adopt a value for the efficiency of the downconversion process ($\varepsilon_\text{DC})$ of $60\,\%$, defined as the percent energy in the QP system after downconversion compared to the original event energy.

    \item \emph{QP trapping--} The remaining QPs will settle to a stable population with energy of $\sim \Delta_\text{island}$ and will diffuse until they either recombine into Cooper pairs over a timescale of $\mathcal{O}(0.1-1)\,\mathrm{ms}$ ~\cite{doi:10.1080/14786436908228699,QP_recomb} (in Al), are trapped by impurities in the film, or ideally get trapped in the lower-$T_c$ superconductor nearby. A QP has an opportunity to become trapped once it diffuses to the overlapping region between the island the LG material, called the `trap'.  When the QP enters the trap, it has an energy of roughly $\Delta_\text{island}$, which is now much larger than the effective gap of the trap at $\Delta_\text{trap}$. These `high energy' QPs now have more available energy states to decay into and a second downconversion process takes place. Once this downconversion process begins in the trap, the average QP energy is now less than $\Delta_\text{island}$ and the QPs can no longer enter the island---effectively confining them to the volume of the trap.
    
    \item \emph{QP multiplication--} In addition to the trapping effect described above, we get an additional benefit from the QP downconversion of the enhancement in QP number. When the high energy QPs downconvert into phonons and lower energy QPs, there is a net energy loss due to a percentage of sub-gap phonons just as was the case in the islands, however as long as the ratio of the two superconducting gaps is greater than three, there is a net increase in the total QP number. Since the SQUAT is sensitive to total QP \emph{number}, this results in a net signal increase. This effect is illustrated by a simple Monte Carlo simulation shown in Fig.~\ref{fig:QP_gain_eff} (the details of which are described in Appendix~\ref{appendix:qp_model}), showing that for a suitable island to trap SC gap ratio, a gain of 5-10 is achievable.

    \item \emph{Multiple measurements of QPs--} Finally, we have an additional multiplication step resulting from the ability to measure the same QP multiple times. Once in the trap, the QPs are confined to a small volume near the Josephson junction. If the tunneling time for the particles is much less than the QP lifetime, the QPs will tunnel back and forth across the junction multiple times until recombination occurs. Each tunneling event results in a measurable signal.
\end{enumerate}



To quantify the efficiency of the above process, we define the QP collection fraction as our metric to optimize our design. The collection fraction, $F_\text{collect}$, is defined as the ratio of the measured number of QP tunneling events across the Josephson junction to the total number of stable QPs remaining after the first downconversion process in the island, i.e. $F_\text{collect}$ quantifies the efficiencies of steps 2-4 above. In defining the collection fraction in this way, we are removing the energy loss from the initial downconversion process ($\varepsilon_\text{DC}$) since this parameter is independent of the design itself. For details on exactly how the modeling for $F_\text{collect}$ is done, the reader is referred to Appendix~\ref{appendix:qp_model}.

A plot of the QP collection fraction as a function of QP trap and qubit island characteristic lengths, $\ell_\text{trap}$ and $\ell_\text{island}$ respectively, for a $500\,\mathrm{nm}$ square JJ can be seen in Fig.~\ref{fig:QP_gain_eff}. This figure highlights the utility of this sensing mechanism; despite the fact that this device adds the complication of the QP tunneling step beyond that of its QET counterpart~\cite{qet}, this efficiency loss can be made up by the addition of the QP multiplication and measurement of multiple tunnel events -- allowing for collection fractions that are greater than unity. 

 \begin{figure}
    \centering

\includegraphics[width=.9\linewidth]{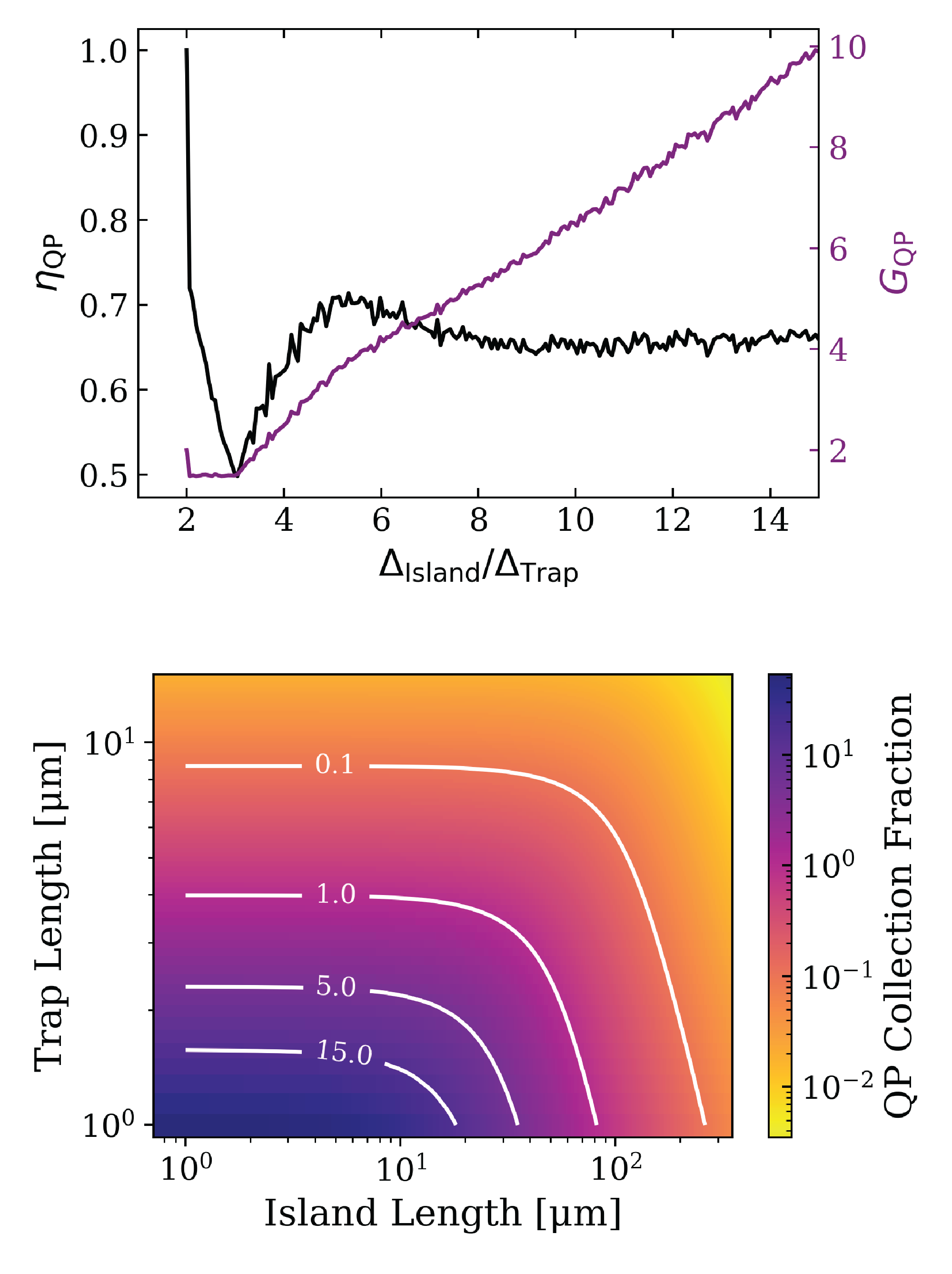}
    \caption{\textbf{Top:} \emph{Left Axis--} Simulated percentage of energy remaining in the QP system ($\eta_\text{QP}$) as a function of incident particle energy relative to the trap gap. In this case the incident particle is a QP of energy $\Delta_\text{Island}$, in units of the trapping material's gap $\Delta_\text{Trap}$. \emph{Right Axis--} Simulated QP number gain, Eq.~\ref{eq:qp_gain}, as a function of SC gap ratio using $\eta_\text{QP}$ from the same figure. \textbf{Bottom:} Modeled QP collection fraction, $F_{\rm collect}$, as a function of trapping region length and island length for a $500\,\mathrm{nm}$ square Josephson junction with $100\,\mathrm{nm}$ thick Al and LG films. The model accounts for multiple QP tunneling events across the junction, which in principle can result in an efficiency greater than unity.}
    \label{fig:QP_gain_eff}
\end{figure}


\textit{Qubit Tuning ---} In order to optimize the readout of these sensors, we must first tune the SQUAT design. Here, we focus on three main design parameters:
\begin{enumerate}
    \item The undressed resonance frequency, $f_0$;
    \item The maximum frequency separation of the even and odd parity states, $2\chi_0$;
    \item The total quality factor of the qubit coupled to the feed line, $Q$; 
\end{enumerate}

\begin{figure}[t]
    \centering
\includegraphics[width=\linewidth]{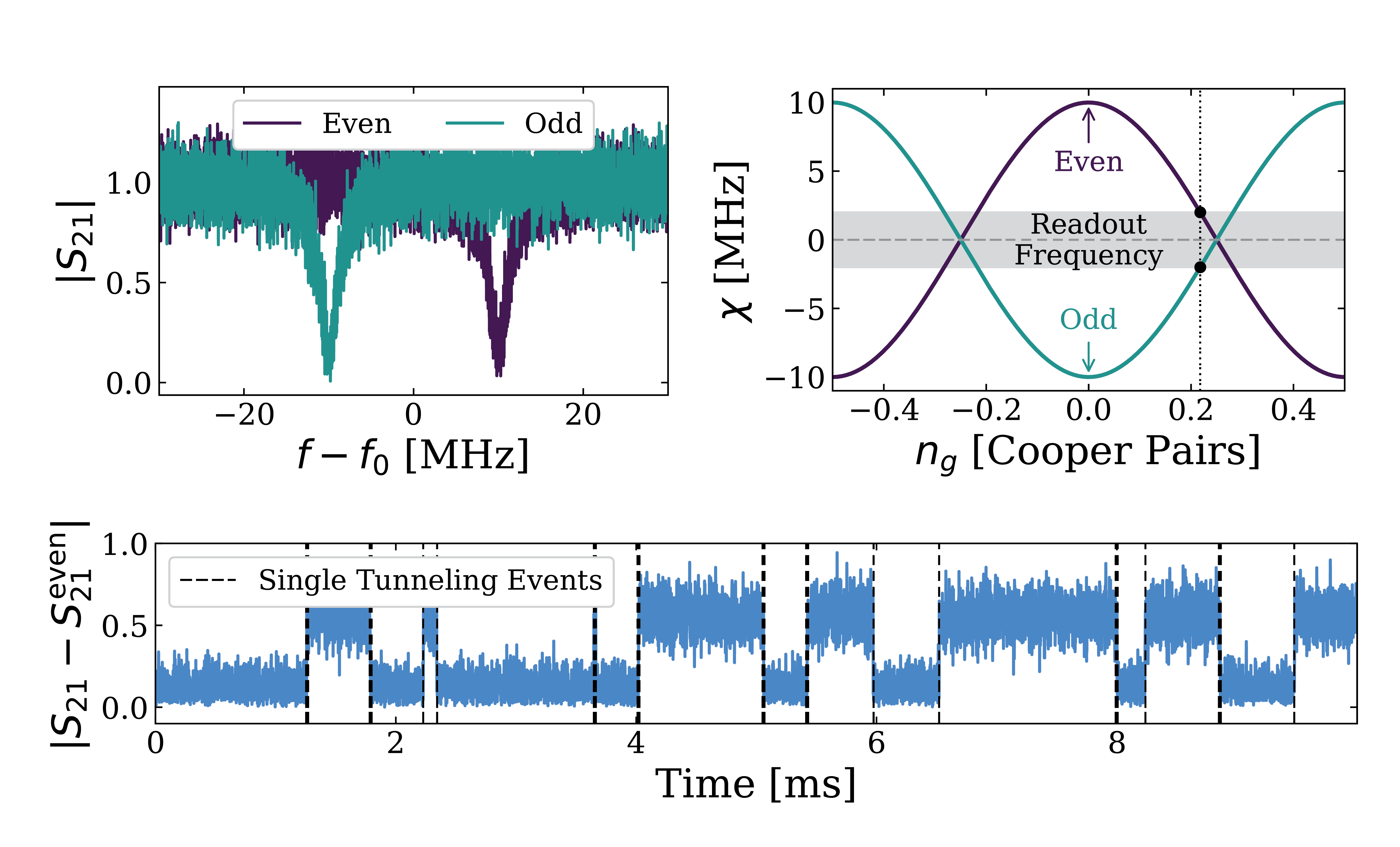}
    \caption{Simulation of the signal readout with $\chi_0=10\,\mathrm{MHz}$ and an effective quality factor of $Q=1000$. \textbf{Left)} Magnitude of $S_{21}$ showing the frequency separation between the odd and even parity states. \textbf{Right)} Charge dispersion of the sensor normalized to the readout frequency, showing the even and odd states. Shown in grey is the target dispersion ($\chi_0$). \textbf{Bottom)} Noisy simulation of single QP tunneling background events with an example dark count rate of 1\,Hz.} 
    \label{fig:readout_signals}
\end{figure}

We model the qubits as notch filters, assuming an ideal transmission line, as \cite{Probst_2015},
\begin{equation}\label{eq:notch}
    S_{21}(f,Q,f_q) \approx 1-\left[1+2iQ\left(\frac{f-f_q}{f_q}\right)\right]^{-1}
\end{equation}
where the qubit resonance frequency $f_q$ is a function of qubit charge $n_g$ and state spacing $2\chi_0$. Converting $E_{01}$ from Eq.~\ref{eq:energy} into frequency, we see that (in the transmon limit), $f_q$ is well approximated by the sinusoidal function
\begin{equation}\label{eq:qfreq}
    f_q(\chi_0,n_g) \approx \chi_0\cos(\pi n_g)+f_0 \,.
\end{equation}
We can choose a charge operating point $n_g$ by applying a charge bias using a dedicated charge gate or along the readout line as was demonstrated in \cite{echternach_2018}.  Note that the bias can either be swept (in the case of biasing on the readout line) or, for individual charge lines, be adjusted during running to account for any slow offset charge fluctuations in the system. This charge bias tunes the state frequency spacing, represented in Eq.~\ref{eq:energy} as $\chi=\chi_0\cos(\pi n_g)$.  When a QP tunneling event imparts a fixed charge shift of 1 electron (shifting the overall phase by $\pi$) the magnitude of the shift of $f_q$ is a fraction of the maximum shift set by the operating point (see Fig.~\ref{fig:readout_signals}, right).

From Eqs.~\ref{eq:f0} and~\ref{eq:dispersion}, we can see that $f_0$ and $\chi_0$ are both determined by $E_J$ and $E_C$, and thus these parameters must be co-optimized. We first choose $f_0$ to be in the operating range of standard RF electronics, e.g. $\sim4-8$\,GHz (this is also informed by available quantum-limited readout).  We must also assure that the state separation (determined solely by $E_J/E_C$) is in the weakly-charge sensitive limit, $\xi\sim\mathcal{O}(1-10)$.  Note that we will return to the precise optimization of $\chi_0$ in the next section.

With $f_0$ and $\chi_0$ chosen, we turn to the question of resonance bandwidth, which is given by $\mathrm{BW}\approx f_0/Q$. The sensors can be readily fabricated to have high-enough internal fidelity that the limiting quality factor is determined by that of the qubit-to-feedline coupling, $Q\approx Q_c$. The optimal bandwidth depends on the readout method, as discussed later in this section and in Appendix~\ref{appendix:resonator}.  In order to effectively resolve phonon events at the timescale of $1\,\mu\mathrm{s}$ we must achieve a bandwidth of $\mathrm{BW}\gtrsim 1\,\mathrm{MHz}$, implying $Q_c \lesssim \mathscr{O}(1000)$.  

Using simulations in HFSS we have demonstrated that generating designs within the optimal parameter space is easily achievable; an example of a SQUAT design tuned for phase readout is shown in Fig.~\ref{fig:sensor_diagram}a.  The fact that an optimization is possible can be made intuitive by thinking about how each parameter depends on the geometry:
\begin{itemize}
    \item $Q\approx Q_c$ depends almost entirely on the capacitive coupling between the qubit and the feedline.  So, tuning $Q$ is achieved by changing the island-feedline separation.
    \item $E_C$ is inversely dependent on the effective capacitance between the islands as set by the full capacitance matrix between each element, including the inter-island capacitance as well as the capacitance to the ground, charge line, and feedline \cite{Schuster_thesis}.  Although the capacitance between the islands and the feedline contributes to this total, it only contributes a fraction of the total capacitance and so can be somewhat decoupled from the optimization of $Q$.
    \item $E_J$ depends on junction parameters including the junction area and thickness.  Although the junction effectively acts as a parallel plate capacitor, this capacitance is also less than the direct island-to-island capacitance that dominates $E_C$.
    \item To choose $\chi_0$, one can fix the $E_J/E_C$ ratio.
    \item Then, with fixed ratio, $E_J$ and $E_C$ can be varied together to get an appropriate $f_0$.
\end{itemize}
The optimal $\chi_0$ and $Q$ are determined by the need to optimize readout signal-to-noise ratio (SNR).  

In order to read out the SQUAT, a signal tone of fixed frequency is pumped through the feedline; we will discuss which frequency is optimal later.  We will focus on a transmission measurement, although a reflection measurement is also possible in principle. Given an input readout tone $V_r$, the signal is described by a complex voltage,
\begin{equation} \label{eq:s21voltage}
    V_{s} = S_{21} V_{r} \,.
\end{equation}
While we measure voltage directly, we convert the voltage signal to power and phase, which allows us to factor out impedance sources in our sensitivity calculations. The measured power transmitted (`signal' power) is thus 
\begin{equation}
    P_{s} = |S_{21}|^2P_r
\end{equation}
with a phase shift of
\begin{equation}
    \tan(\theta) = \frac{\mathrm{Im}(S_{21})}{\mathrm{Re}(S_{21})},
\end{equation}
where $P_r$ is the input tone power. This also allows us to determine the power dissipated in the device, with a precision determined by the degree to which attenuation and gain in our readout system are known.

There are two fundamental modes in which we can read out the SQUATs in this two-dimensional basis. The first mode is to observe a shift in the amplitude of the transmitted signal, with phase shift fixed at zero; the second is to observe a shift in the signal phase with amplitude unchanged. Both correspond to states separated by the diameter of the resonator circle in the IQ space, but amplitude readout is less sensitive to the precise frequency dispersion of the qubit, while properly-optimized phase readout is less sensitive to readout frequency. These cases are discussed in more detail in Appendix~\ref{appendix:resonator}, and summarized qualitatively here to focus on readout optimization results for the SQUAT.

In the amplitude readout mode, the optimal signal would be one in which $S_{21}$ equals 1 or 0 for the different parity states, so that the signal tone completely appears ($V_s=V_r$) or disappears ($V_s=0$) during a tunneling event.  This situation can be achieved if the readout tone is within the bandwidth of one of the parity states and there is effectively no overlap of the two states, e.g. the state separation is much larger than the resonator width:
\begin{equation}
    2\chi\gg f_0/Q \,.
\end{equation}

In the second readout mode, in which we wish to measure distance between states in the complex plane, a different optimization of $Q$ and $\chi$ is necessary. Consider placing the readout tone at a frequency between the parity state frequencies, such that the transmitted magnitude is constant, but a relative phase shift is observed. As we derive in detail in Appendix~\ref{appendix:resonator}, the measured phase shift is maximized when
\begin{equation}
    2\chi\sim f_0/Q \,.
\end{equation}
The main advantage of this second readout scheme is, as shown in the appendix, that a sensor optimized in this way will produce a fixed shift in the IQ plane for any readout frequency between the two parity states. This means we don't need to worry about precise readout frequency optimization. In this case, in which noise is uncorrelated in phase, we will not see a large difference in readout fidelity as a function of readout frequency. For noise primarily in the phase or amplitude direction, we can adjust readout frequency to minimize the amount of noise projected onto our readout vector.

For both of these cases, we show in Appendix~\ref{appendix:fidelity} that the qubit fidelity is just defined by the noise temperature and readout power according to the equation
\begin{equation}\label{eq:SNRAmp}
    SNR^{-1} = \sigma^{2} = \frac{2P_{n}}{\epsilon_{r}P_{r}} = \frac{2k_B T_n f_{bw}\eta\left(\frac{hf_0}{k_BT}\right)}{P_{r}\left[1-\exp\left(-\frac{\hbar f_0^2}{Q_cP_{r}}\right)\right]} \,.
\end{equation}
where $f_{bw}$ is the readout bandwidth (inverse integration time), $T_n$ is the system noise temperature, $f_0$ is the readout frequency of the SQUAT, and $\sigma^2$ is the variance. Readout efficiency, $\epsilon_r$, and quantum noise corrections, $\eta$, are described in Appendix~\ref{appendix:fidelity}. For $T_n\sim$1.8 K and $f_0\sim$6 GHz, Eq.~\ref{eq:SNRAmp} dictates that we can achieve high readout fidelity above -135\,dBm readout power for an integration time corresponding to $f_{bw}=10^4\,\mathrm{Hz}$. With a quantum-limited readout, we find the reduced equation
\begin{equation}\label{eq:SNRAmpReduced}
    SNR = \frac{P_{r}\epsilon_r}{2 h f_0 f_{bw}} \,.
\end{equation}
where high fidelity is achieved above $P_r\approx-140\,\textrm{dBm}$ at $f_{bw}\sim10^{5}$ Hz. These readout powers have already been used to demonstrate high readout fidelity for state-of-the-art qubits. Real devices will need to map out the tradeoffs between parasitic pair-breaking of the readout tone and noise-limited readout, and may benefit from stimulated emission from the SQUAT at higher readout powers.

In practice, the difference in SNR between phase and amplitude readout will be determined by the level of phase noise in the system (which we elaborate on in Appendix~\ref{appendix:phase_noise}), and will be assessed during first device tests.  We have the freedom to tune state separation to optimize for either readout mode.


\begin{figure*}
\centering
\includegraphics[width=\linewidth]{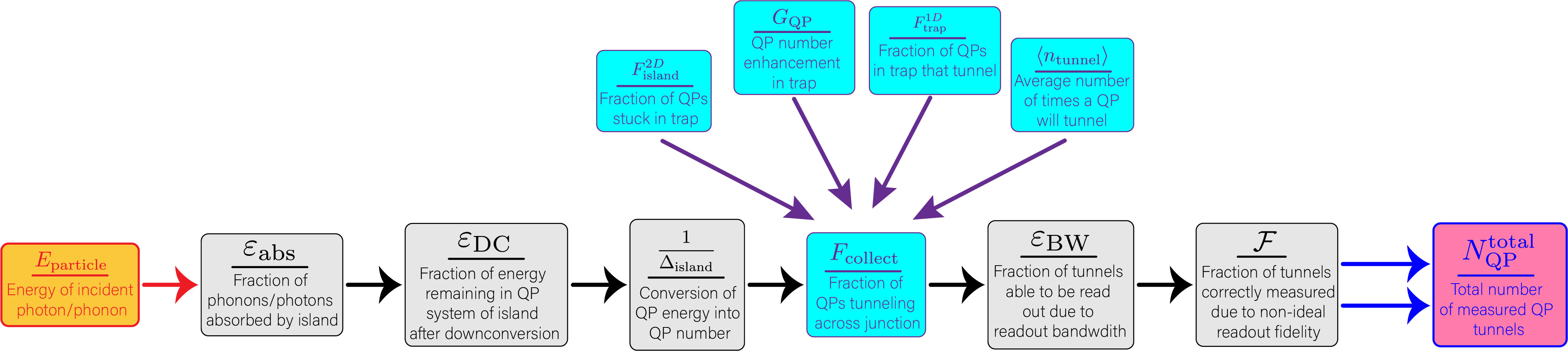}
\caption{Schematic flow chart showing all the efficiency and conversion factors going from incident particle energy ($E_\text{particle}$) to total number of measured QPs ($N_\text{QP}^\text{total}$). Each efficiency factor is defined as fraction of the measured quantity that is output by the process in consideration, divided by the amount that is left over from the previous process, e.g. output vs input from left to right. In this way, all factors can be multiplied. The definitions for each efficiency can be found: $\varepsilon_\text{abs}$: Eq.~\ref{eq:phonon_eff} (for the phonon case), $\varepsilon_\text{DC}=60\%$, $F_\text{collect}$: Eq.~\ref{eq:total_QP_collect} (with definitions there within), $\varepsilon_\text{BW}$: Eq.~\ref{eq:phonon_BW} phonon (Eq.~\ref{eq:BW_eff_eval}) (photon), and $\mathcal{F}$: Eq.~\ref{eq:fidelity}.}
\label{fig:eff_chart}
\end{figure*}



\textit{Estimated Sensor Performance---}We can combine all of our known efficiency factors to predict the final energy sensitivity of the device to both direct absorption of photons and measurement of athermal phonons in the substrate. Recall that we are measuring total QP number $n_\text{qp} \approx \eta_\text{qp}/\Delta_\text{island}$, where $\eta_\text{qp}$ is the total energy in the QP system. The energy in the QP system can be written as the product of the impinging particle energy $E_\text{particle}$ (either from a photon or phonon), the absorption efficiency ($\varepsilon_\text{abs}$), the efficiency from the downconversion process ($\varepsilon_\text{DC}$), and remaining efficiency factors described below. Thus, our measured number of QP tunneling events ($N_\text{QP}^\text{total}$) will be

\begin{equation}
    N_\text{QP}^\text{total} = E_\text{particle}\times \varepsilon_\text{abs} \times \varepsilon_\text{DC} \times \frac{1}{\Delta_\text{Island}}\times F_\text{collect} \times \varepsilon_\text{BW} \times \mathcal{F},
    \label{eq:number_qp}
\end{equation}
where $F_\text{collect}$ is the QP collection fraction given by Eq.~\ref{eq:total_QP_collect}, $\varepsilon_\text{abs}$ is the efficiency of the particle absorption in the sensor, i.e. the antenna coupling for photons or the signal loss due to passive surfaces for the phonon case, $\varepsilon_\text{BW}$ is the measurement efficiency due to the finite sensor bandwidth from Eq.~\ref{eq:phonon_BW}, and $\mathcal{F}$ is the readout fidelity given by Eq.~\ref{eq:fidelity} (See also Fig.~\ref{fig:eff_chart} for a pictorial flow of how the output signal is calculated). Considering only the low energy threshold and operating with a readout bandwidth of $\sim 10\,\mathrm{kHz}$ with standard RF readout hardware, we can ignore dynamic range effects from the finite bandwidth and set $\varepsilon_\text{BW}=1$ and $\mathcal{F}=1$ (see Appendix~\ref{appendix:BW} and Appendix~\ref{appendix:fidelity} for more details on the effects of bandwidth on the dynamic range and readout fidelity of the SQUAT.)

Considering first the athermal phonon detection, taking the worst case scenario that all the phonons are above the Nb gap, our phonon collection efficiency due to passive surface losses is roughly $\varepsilon_\text{abs}\approx 37\,\%$. Note that this value is relatively independent of individual sensor size, and the total sensor number can be changed to keep the surface coverage the same. Assuming a readout mode such that single QPs can be counted, Eq.~\ref{eq:number_qp} can be converted into an effective phonon energy resolution by setting $N_\text{QP}^\text{total} = 1$ and solving for $E_\text{particle}$. This can then be solved for a variety of sensor geometry parameters by using Eq.~\ref{eq:phonon_eff}, Eq.~\ref{eq:total_QP_collect}, and $\varepsilon_\text{DC} = 60\,\%$, with the parameters listed in Appendix~\ref{appendix:qp_model}.

In Fig.~\ref{fig:energy_res} we show the expected athermal phonon energy sensitivity threshold as a function of sensor island and trap length (see Fig.~\ref{fig:sensor_diagram} for definitions) instrumented on a $500\,\mu\mathrm{m}$ thick Si wafer. From the figure, we can see that for various geometries, the SQUAT can easily achieve single meV phonon thresholds. This represents the underlying fundamental sensitivity for the SQUAT, but is in reality the lower bound of detection threshold. In practice, the threshold will ultimately be limited by the dark rate from excess non-equilibrium QPs tunneling across the junction. While dark rates will always be a limitation for counting based sensors, transmon qubits have achieved dark rates as low as $\mathcal{O}(10-100)\,\mathrm{mHz}$~\cite{yelton2024modeling}, sufficiently low enough to enable us to operate in the single to few QP counting regime.

The photon energy sensitivity is expected to be similar to the phonon scenario since $\varepsilon_\text{abs}$ is very close for both cases. A full simulation of the antenna structure is left for future work, but from initial conservative estimates we expect that sensitivity to single sub-THz photons is reasonably achievable. A simulation of the expected sensor response as a function of photon energy and island length is shown in Fig.~\ref{fig:energy_res}.

In order to illustrate the substantial improvement in sensitivity that the QP trapping stage adds, we have plotted the simulated QP collection efficiency of the SQUAT sensor with a $4\,\mu\mathrm{m}$ trap compared to an equivalent geometry sensor but without the trap in Fig.~\ref{fig:energy_res} right. As can be seen, the trapping region adds multiple orders of magnitude of phonon or photon sensitivity to the detectors.

While we focus in this section primarily on the low energy threshold of the SQUAT, the dynamic range of the measurement can span multiple orders of magnitude in energy with minimal signal loss and is a function of the measurement readout bandwidth, discussed in Appendix~\ref{appendix:BW}.


\textit{Discussion---}In this letter we have presented the design sketch and modeling of a novel detector concept based on superconducting qubit technology, which if properly optimized, would allow for sensitivities to particle interactions at the meV scale. While the majority of our QP model we have developed in Appendix~\ref{appendix:qp_model} is based on previously measured data, we have made a few (conservative) assumptions, which we plan to measure in future work. Over the next few years, our R\&D path towards realizing these devices is to:
\begin{itemize}
    \item Fabricate devices of the geometry presented in this paper using only Al for the islands and junctions (no QP trapping) to understand the optimization of the readout and qubit parameters.
    \item Optimize the fabrication of junctions on low-gap materials (e.g. Ti, Hf, and AlMn) to understand the QP tunneling probability of the junctions.
    \item Fabricate the complete SQUAT sensor as described in this paper to measure the QP trapping probability at the Al/LG overlap.
\end{itemize}
With the LG trapping and tunneling parameters measured, a further optimization of the sensors will be possible. Once the sensitivity is maximized, the SQUATs have a large number of use cases, primarily in the low-energy, rare-event regime. The sensor offers two different pathways for event detection:
\begin{enumerate}
    \item The antenna structure of the sensor itself allows this device to be sensitive to single THz photons, allowing the devices to used in searches for QCD axion DM such as BREAD \cite{Liu2022}. Additionally, this device is sensitive to electromagnetically-interacting fermionic light dark matter (LDM) via direct absorption or scattering in the island itself as proposed in~\cite{das2022dark}.
    \item The sensitivity to athermal phonons in the device makes it well-suited to detect LDM at the $\mathcal{O}(\mathrm{MeV})$ mass scale and sub-eV bosonic DM such as dark photons. Absorption of LDM and far-IR photons is possible in Si via a multiphonon excitation process~\cite{two_phonon, Hochberg_ldm_simiconductors}. The LDM or photon coupling strength can be increased by using a polar crystal substrate~\cite{PhysRevD.101.055004, Trickle_2020} or a low-bandgap semiconductor~\cite{rosa2020colossal, Piva_2021}. In all of these cases, sensitivity to acoustic phonons at the $\mathcal{O}(1-10\,\mathrm{meV})$ scale is required.
\end{enumerate}

\begin{figure*}[t]
\centering
\includegraphics[width=\linewidth]{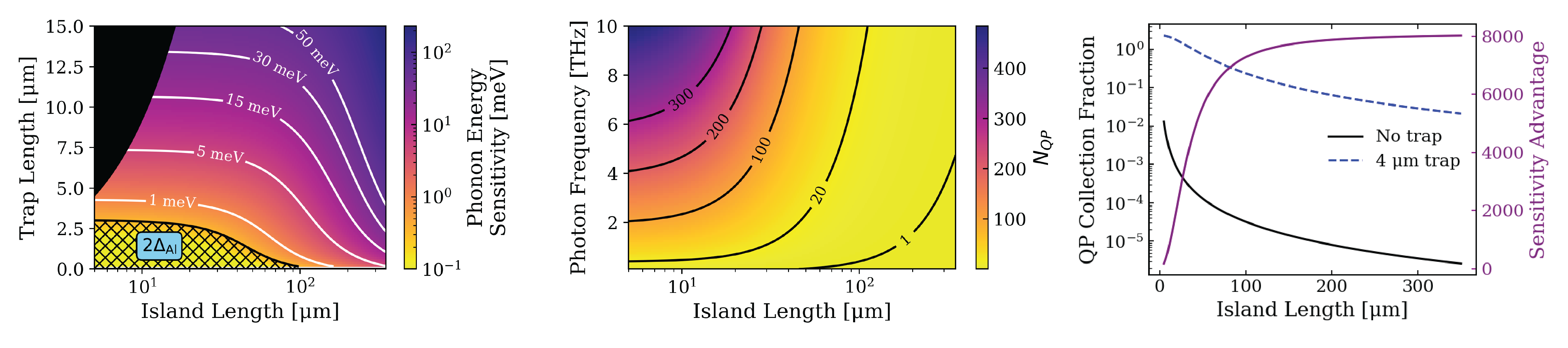}
\caption{\textbf{Left} Simulated phonon energy resolution, defined as the phonon energy which would cause a single parity flip in the SQUAT, given by Eq.~\ref{eq:number_qp}, as a function of island and trap length (see Fig.~\ref{fig:sensor_diagram}) assuming negligible change in fill factor as a function of island length for an array of SQUATs instrumented on a $500\,\mu\mathrm{m}$ thick wafer. Shown in the hatched region is twice the superconducting gap of Al, and the blacked out regions represent non-physical design parameters. \textbf{Middle} Simulated QP tunneling number $N_\mathrm{QP}$ across the junction as a function of incident photon frequency and island length for a $4\,\mu\mathrm{m}$ trap. Note, we are not factoring in the antenna effect when changing the island length at this point -- simply assuming that a $\sim\,50\%$ photon absorption efficiency can be measured. \textbf{Right} Comparison of the SQUAT with a $4\,\mu\mathrm{m}$ trap vs a device of the same geometry, but with no trapping region. Shown in purple on the right axis is the sensitivity advantage of including QP trapping.}
\label{fig:energy_res}
\end{figure*}

While there already exists a large amount of competition in the space of superconducting sensors for low-energy rare-event measurement~\cite{fink_TES, doi:10.1063/1.1596723, 918addfc209841dca7f245dfe9288a30, Goldie, INRIM, spica, fink_PD2}, the device we have proposed potentially offers multiple orders of magnitude of improvement in terms of absolute energy sensitivity beyond these currently existing detectors. Furthermore, because each qubit sensor is naturally read out individually, they inherently have a powerful background discrimination tool by determining the location in the detector in which the event occurred.

\textit{Acknowledgements---}Thanks to Sasha Anferov for several useful conversations about qubit modeling. This work was supported in part by the US Department of Energy Early Career Research Program (ECRP) under FWP 100872. Caleb Fink is supported by the  Laboratory Directed Research and Development program of Los Alamos National Laboratory. C. P. Salemi is supported by the Kavli Institute for Particle Astrophysics and Cosmology Porat Fellowship. This work was supported in part by the DOE Office of Science High Energy Physics QuantISED program.



\providecommand{\noopsort}[1]{}\providecommand{\singleletter}[1]{#1}%
%


\appendix

\section{Resonator Readout Modes}\label{appendix:resonator}

When measuring time-domain SQUAT behavior in a highly multiplexed environment, we are unlikely to be able to retain both quadrature components, and we want to see how streaming only a single quadrature will impact our ability to discriminate between parity states. This is not the optimal readout mode, as retaining both quadratures allows for optimal signal to noise as discussed in the next appendix. This appendix serves instead to illustrate how we should optimize dispersion such that these two limited cases can have maximal sensitivity to absorbed energy.

We can operate SQUATs in two distinct modes: amplitude or phase readout. For this discussion, recall that our voltage signal is the transmission signal of a complex pulsed readout tone $V_r$:
\begin{equation}
    V_s = S_{21}V_r.
\end{equation}

\begin{figure}[h]
    \centering
\includegraphics[width=\linewidth]{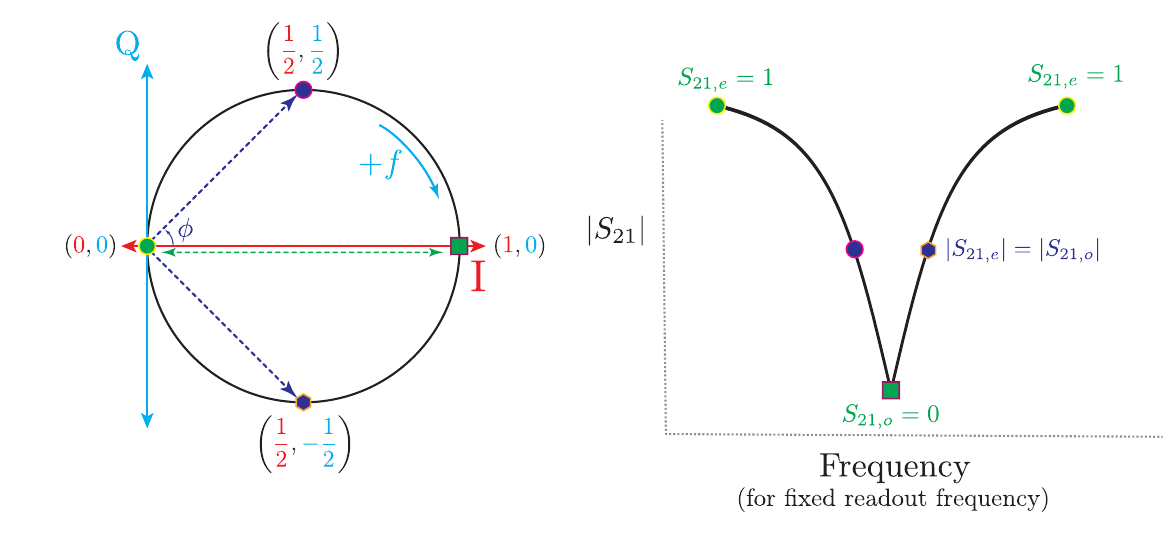}
    \caption{\textbf{Left:} Schematic of the two qubit readout scenarios shown on the IQ circle. The Green dots represent the case of amplitude readout, in which the state separation is much larger than the bandwidth of the readout such that $S_{21}$ can take on the values of 1 or 0. The purple dots represent the case of phase readout, in which the amplitude of $S_{21}$ is equal for the odd and even parity states but the phase difference between the states is maximized. \textbf{Right:} The corresponding points shown in frequency space, keeping the readout frequency constant.}
    \label{fig:IQ}
\end{figure}

\textit{Amplitude Readout---}Since $|S_{21}|$ can only take on values between zero and one, the maximum and minimum amplitudes of the signal voltage are $|V_s| = |V_r|$ and $|V_s| = 0$ respectively. Optimal amplitude readout is accomplished in practice by designing a state separation $2\chi$ that is much larger than the bandwidth of the SQUAT ($2\chi \gg f_0/Q$). Thus the readout frequency can be set at one parity state while the other state is outside the bandwidth of the readout. This can be seen visually on the IQ circle in Fig.~\ref{fig:IQ}. This is the readout mode we would be forced to use if only one state lies on the resonator circle, which is the case for the condition $2\chi \gg f_0/Q$. We would only have maximal contrast between states for such a device with a tone placed directly on resonance with the SQUAT, which would represent maximal energy coupling into the device.

\textit{Phase Readout---}Consider instead if we were to optimize the SQUAT dispersion such that we can maximize state contrast in the phase of the transmitted signal tone. Microphyiscally, this corresponds to photons scattering off of the SQUAT and picking up a state-dependent phase shift. To solve for the optimal dispersion given a known SQUAT quality factor, we first simplify the problem such that we remove all information from the amplitude readout axis. This corresponds to the condition
\begin{equation}
   |S_{21,e}| = |S_{21,o}|.
\end{equation}
Solving this equality thus gives us an optimization in which both amplitude and phase readouts contain maximal parity information. If this is the case, as we will show in the next appendix, we can guarantee that full quadrature readout at any frequency between the two parity states has an equal amount of discrimination, which will be important for optimizing sensor fidelity.

Let us first recall the definition of $S_{21}$ when limited by the quality factor of the coupling
\begin{equation}\label{eq_s:notch}
    S_{21}(f,Q,f_q) \approx 1-\left[1+2iQ_c\left(\frac{f-f_q}{f_q}\right)\right]^{-1}
\end{equation}
where the qubit resonance frequency is given by 
\begin{equation}
    f_q(\chi_0,n_g) \approx \chi+f_0 
\end{equation}
recalling that the effective dispersion $\chi=\chi_0\cos(\pi n_g)$. For simplicity, let us write $S_{21}$ as 
\begin{align}
    S_{21} &= \frac{-i \alpha}{1-i \alpha}\\
    &=\frac{\alpha^2 - i \alpha}{1+\alpha^2}
\end{align}
where $\alpha$ can be written for the even and odd states as
\begin{equation}
    \alpha_{e/o} = 2 Q\left(\frac{\pm\chi}{f_0 \pm \chi}\right).
\end{equation}
Separating $S_{21}$ into IQ space we get that
\begin{align}
    \text{I} \equiv \operatorname{Re}(S_{21}) &= \frac{\alpha^2}{1+\alpha^2}\\
    \text{Q} \equiv \operatorname{Im}(S_{21}) &= \frac{-\alpha}{1+\alpha^2}
\end{align}
and the phase angle is 
\begin{equation}
    \tan(\phi) = {\frac{\text{Q}}{\text{I}}}
\end{equation}
Considering two points on the IQ unit circle (see Fig.~\ref{fig:IQ}), the maximum separation (maximum signal) of phase between the points occurs when they are separated by a phase shift of $\pi/2$. Any solution with this phase shift will be valid, but this yields the simplest solution in the case of the phase readout described earlier, where the states are symmetric about the I axis.

For I symmetric states, the maximum separation corresponds to $\tan(\pm\pi/4) = \pm 1$.
\begin{align}
       \pm 1 &= \frac{\text{Q}}{\text{I}} \\
       \pm 1 &= -\frac{1}{\alpha}\\
       \pm 1 &= \frac{f_0 \pm\chi}{2Q_c(\pm \chi)}\\
       1 &\approx \frac{f_0}{2Q_c\chi}
\end{align}
since $f_0 \gg \chi$. Finally, this gives us the optimum dispersion for a given bandwidth, 
\begin{equation}
    2\chi \approx \frac{f_0}{Q_c}.
\end{equation}
which is essentially a condition for the minimum dispersion we need to design into the SQUAT to ensure we can always tune the charge bias to find an optimally separated set of parity states. In the case where we expect significant charge noise, there is some value in designing a more fine-tuned dispersion as close to this value as possible; we will discuss this in the context of non-ideal noise sources in the next appendix.

As a final note, we point out that $Q_{c}$ is not itself an intrinsic quantity, but instead is defined as $Q_c = \frac{f_0}{\Gamma_{d}}$, where $\Gamma_d$ is the decay rate of photons from the SQUAT back into the environment. We thus find that this condition is just $2\chi = \Gamma_d$, or in other words, we want to ensure that the SQUAT decays quickly enough that we can distinguish between parity bands. This decay rate is also essentially the maximum interrogation rate of the SQUAT, and so tradeoffs in readout bandwidth dictate limits on the allowable dispersion of the SQUAT.

\section{Readout Fidelity}\label{appendix:fidelity}

Given the discussion of optimizing SQUAT dispersion vs bandwidth in Appendix~\ref{appendix:resonator}, we are prepared now to talk about readout fidelity for this type of sensor design. We will show, as is typical of readout in I,Q space, that proper optimization of the design of the sensor will reduce sensitivity to readout frequency within a range of the nominal central frequency determined by $Q$. For those familiar with this general optimization, proceed to Equation~\ref{eq:photonSaturation} for the discussion of how photon saturation affects readout optimization.

\paragraph{Signal Magnitude} The readout in either of the limiting cases discussed in the previous section is a measurement of transmitted power across the resonator given a known input tone power and phase. Alternatively, we can frame the signal amplitude as the distance between even and odd parity states in IQ space. In the general case, the distance between two points in IQ space is 
\begin{equation}
    |\Delta S_{21}|^2 = (I_1 - I_0)^2 + (Q_1 - Q_0)^2
\end{equation}
For the amplitude case, change in measured power transmitted through the qubit, $\Delta P_s$, for an input power, $P_r$, is 
\begin{equation}
    \Delta P_s = |\Delta S_{21}|^2P_r = \left[(1 - 0)^2 + (0 - 0)^2\right]P_r = P_r \,.
\end{equation}
For phase readout, where the signal is constant in I and only the sign of Q changes, we find
\begin{equation}
    \Delta P_s = |\Delta S_{21}|^2P_r = \left[\left(\frac{1}{2} - \frac{1}{2} \right)^2 + \left(\frac{1}{2}  + \frac{1}{2}\right)^2\right]P_r = P_r \,.\end{equation}
Thus in terms of signal amplitude, there's no difference between these readout modalities. 

Let's now consider the general case for the optimization of the phase readout (in which we assume that we have access to both the phase and amplitude of $S_{21}$). Tuning the SQUAT to a point where $2\chi \approx \frac{f_0}{Q_c}$ ensures, as described earlier, that the even and odd states lie opposite each other on the resonance circle. For a rotation angle of $\theta$ about the point (I,Q) = (1/2,0), we find that
\begin{align}
    I &= \frac{1}{2} \cos(\theta+\Omega) + \frac{1}{2} \\
    Q &= \frac{1}{2} \sin(\theta+\Omega)
\end{align}
where $\Omega = \pm \pi/2$ for the even/odd state. We thus find that
\begin{align}
    \Delta I &= \frac{1}{2}\left[\left(\cos\left(\theta+\frac{\pi}{2}\right) + 1\right) - \left(\cos\left(\theta-\frac{\pi}{2}\right) + 1\right)\right] \\
     &= \frac{1}{2}\left[-\sin(\theta) - \sin(\theta)\right] = -\sin(\theta) \\
     \Delta Q &= \frac{1}{2}\left[\sin\left(\theta+\frac{\pi}{2}\right) - \sin\left(\theta-\frac{\pi}{2}\right)\right] \\
     &= \frac{1}{2}\left[\cos(\theta) - (-\cos(\theta))\right] = \cos(\theta)
\end{align}
which tells us that for the general case of a frequency shift from the optimal readout frequency, for an optimized design,
\begin{align}
    |\Delta S_{21}|^2 &= (\Delta I)^2 + (\Delta Q)^2 \\
    &= \sin(\theta)^2 + \cos(\theta)^2 = 1
\end{align}
Thus the design tuned to this optimal dispersion will have the same signal amplitude regardless of the readout frequency, as the two points will be shifted an equal amount around the resonator circle but maintain unit distance between them. Contrast this with amplitude readout in the non-optimized case, where maximum signal is only achieved when the readout frequency is exactly the frequency of one of the parity states.

\paragraph{SQUAT Saturation} One important caveat to this argument is that, unlike a lumped-element resonator, the SQUAT cannot interact with more than one photon at a time. We can account for this by adding a readout efficiency term to the signal part of our sensitivity calculation, which accounts for the photons that are unable to interact with the SQUAT after the initial photon has been absorbed. This readout efficiency is given by the equation
\begin{equation}\label{eq:photonSaturation}
    \epsilon_{r} = 1-\exp\left(-\frac{\hbar f_0^2}{Q_cP_{r}}\right) = 1-\exp\left(-\frac{\Gamma_{d}}{\Gamma_{\gamma}}\right)
\end{equation}
The frequency dependence comes both from calculating the decay rate for a given coupling factor $Q_c$ and from converting readout power into photon number. 

If we convert power to photon interrogation rate $\Gamma_{\gamma}$ and substitute in our decay rate from earlier (second half of Eq.\ref{eq:photonSaturation}), this becomes clearer. The readout efficiency improves when the interrogation rate is slower than the decay rate. Increasing the decay rate allows for the SQUAT to be readout more quickly, but comes with the side-effect of also coupling it more strongly to the environment. As discussed earlier, increasing decay rate also increases the minimum state dispersion, reducing the multiplexing factor. Finally, Eq.~\ref{eq:photonSaturation} also shows that for high readout power we suffer from the limited number of photons interacting with the SQUAT.

\paragraph{Readout Fidelity} Having established that our signal amplitude is always equal to our readout power times readout efficiency $P_{s}=\epsilon_r P_r$, we now consider the noise contribution to our signal in order to determine readout fidelity. Assuming we are limited by Johnson-Nyquist noise (which is uncorrelated to the signal), we find a noise power in an impedance-matched network of 
\begin{equation}
    P_{n} = k_B T f_{bw}\eta\left(\frac{hf}{k_BT}\right)
\end{equation}
where
\begin{equation}
    \eta(x) = \frac{x}{\exp(x)-1}+\frac{x}{2}\,.
\end{equation}
$f_{bw}$ is the readout bandwidth of the measurement and $T$ is the physical temperature of the system. The role of $\eta$ is to introduce quantum corrections from vacuum fluctuations that break the strict temperature dependence of the noise; for low-enough physical system temperature, this is reduced to the simpler, quantum-limited expression
\begin{equation}
    P_{n} \approx f_{bw}\frac{hf}{2}
\end{equation}

If we assume that the noise and signal are uncorrelated, we can continue to use readout power as our signal to write down the signal-to-noise ratio, or equivalently the inverse normalized signal variance, as
\begin{equation}
    SNR^{-1} = \sigma^{2} = \frac{2P_{n}}{\epsilon_{r}P_{r}} = \frac{2k_B T f_{bw}\eta\left(\frac{hf}{k_BT}\right)}{P_{r}\left[1-\exp\left(-\frac{\hbar f_0^2}{Q_cP_{r}}\right)\right]}.
    \label{eq:SNR_ap}
\end{equation}
Here the factor of 2 comes from the fact that we have defined our signal as a difference; we thus need two measurements to establish that difference, and we pick up two noise contributions as a result. Note that given our signal in this case is a difference, the power noise can thus can be treated as a variance; this is different from an absolute measurement where noise power is a strictly positive number.  As discussed in the main body of the paper, correlated noise such as TLS noise will modify the SNR, effectively by making the noise level dependent on the operating point on the IQ circle.

\begin{figure}[h]
    \centering
\includegraphics[width=\linewidth]{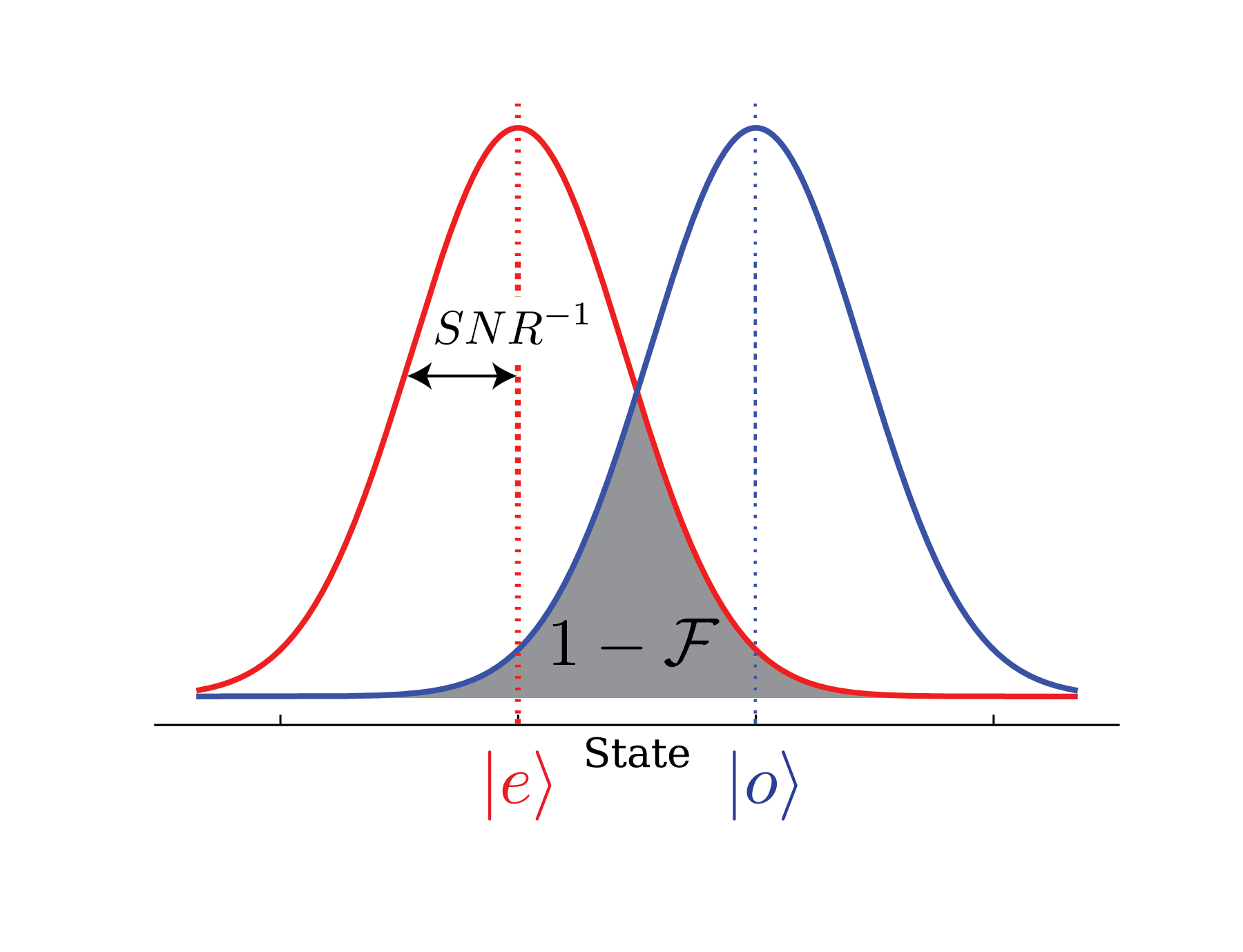}
    \caption{Schematic of readout fidelity calculation.}
    \label{fig:fidelity_diag}
\end{figure}

\begin{figure}[h]
    \centering
\includegraphics[width=\linewidth]{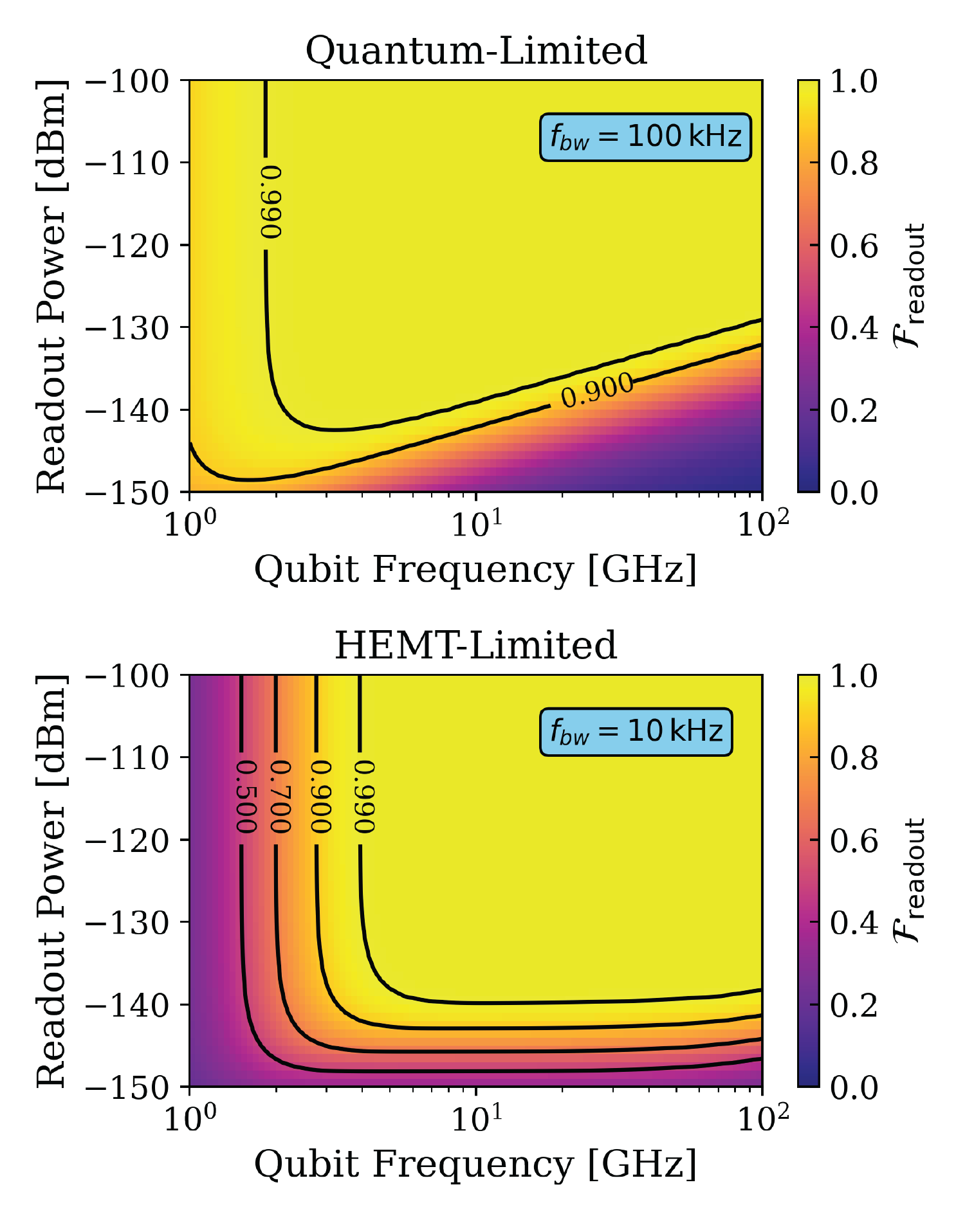}
    \caption{Expected readout fidelity ($\mathcal{F}_\text{readout}$) of sensor as a function of readout power and resonance frequency for a quantum limited readout (Top) and an HEMT limited readout, assuming a noise temperature of $T_n=2\,\mathrm{K}$ (Bottom).}
    \label{fig:fidelity}
\end{figure}

We can now consider the fidelity $\mathcal{F}$ for our readout. Assuming random non-stationary noise, we can calculate the fidelity utilizing the square root of the overlap of two normal distributions (see Fig.~\ref{fig:fidelity_diag}) with $\mu=0,1$ and variance defined in Eq.~\ref{eq:SNR_ap},
\begin{equation}
    \mathcal{F} = 1 - \exp\left(-\frac{SNR}{4}\right).
    \label{eq:fidelity}
\end{equation}
This fidelity measure is the joint probability that the two states overlap, and goes to zero for low SNR, 1 for high SNR. Repeated readouts increase SNR, and thus in general, reduced readout bandwidth can be used to increase fidelity. A plot of the readout fidelity as a function of readout power and qubit resonance frequency for both a quantum limited and HEMT limited readout can be seen in Fig.~\ref{fig:fidelity}.

\subsection{Sources of Phase Noise}
\label{appendix:phase_noise}

In the main body of the text and in this appendix (Appendix~\ref{appendix:fidelity}), we have assumed the ideal case of perfect qubit stability and stationary, uncorrelated noise additive with I and Q. In reality, there are two major noise sources we have to worry about: charge noise and two-level system (TLS) noise. Charge noise affects the point on the dispersion curve and thus the state separation (we can think of this as a differential-mode phase noise). TLS noise affects the mean frequency of the device; it can be thought of as a common-mode phase noise.

In the case of charge noise, we can use the gate capacitance to estimate the effect of a known gate voltage noise on the system; TLS noise will be determined by the occupation and coupling of, e.g., oxide states near the qubit to the qubit islands. As mentioned earlier, the impact of charge noise is affected by the maximum qubit dispersion, as in the frequency-insensitive readout mode we will have a larger charge noise if the dispersion needs to be tuned down to readout the optimal readout point. This may mean that, in a high charge noise environment, we will have to resort to simpler magnitude readout and accept any resulting additional noise from frequency instability of the tone.

\subsection{Comparison to Alternative Architectures}

One obvious difference between the SQUAT and other potential architectures, including the strongly-coupled transmon and the QCD, is the saturation in the qubit response with power leading to diminishing returns as pump power is increased. For quantum-limited readout, this still allows for real-time bandwidth of close to 1 MHz depending on the required state fidelity, but places stringent requirements on the readout hardware and signal filtering required to produce an optimal readout. Here we briefly consider the pros and cons of the SQUAT as compared to alternative architectures.

Consider first retaining the readout resonator so the SQUAT acts essentially the same as a standard weakly charge-sensitive transmon. In this case the coupling is weak, and the coherence time will be long. A Ramsey sequence of X/2, idle, Y/2 can be used to determine charge parity state, and this sequence can be optimized to occur quickly enough for high-bandwidth readout given a system with high single-shot fidelity\cite{Wilen_2021}. For a given initial state, if no parity switch occurs, the final state is mapped back to the initial state. This is practically just limited by the bandwidth of the qubit-resonator system and the speed of the readout electronics, but the procedure to determine Ramsey drive and readout sequences is much more complex, and scaling this method to many resonators is expensive, both computationally and in terms of hardware required. In comparison, the SQUAT can be run in a CW-mode with a much lower cost VNA (only a single drive is required) and multiplexing can be done with continuous wave sources in a manner similar to readout of MKIDs\cite{temples2024performance} or RF-SQUIDs \cite{Yu_2023}. The long-term readout plan is to integrate this continuous wave readout with slow feedback to stabilize offset charge, reducing long-term drift and identifying substrate events from background radioactivity and cosmogenics via correlations in offset charge feedback.

The advantages of the SQUAT extend beyond reduced readout complexity. Where typical charge-sensitive transmons generally require large lumped-element or coplanar waveguide (CPW) resonators for readout, the entire area of the SQUAT is active for sensing. This means that, for phonon sensing, there is not a large loss of phonons into the much larger (by volume) readout resonator, and higher fractional active sensor areas can be produced. For photon sensing, it is possible to make a dense SQUAT array comparable to large area SNSPDs or MKIDs, which is in contrast to the very small active area of QCDs. For applications limited by quantum efficiency, allowing for direct imaging of a photon field with a SQUAT array, such as will be required for BREAD \cite{bread} or photon-limited applications in astrophysics.

\section{Quasiparticle Diffusion Model}
\label{appendix:qp_model}
This diffusion model describes the dynamics of a non-equilibrium QP population in the qubit islands and how they are measured. As described in the text above, this QP population arises from an impinging particle, e.g. a phonon or photon, with energy greater than twice the superconducting bandgap of the island. The incident particle breaks a Cooper pair into highly energetic QPs, which then undergo a downconversion process into phonons and lower energy QPs until a quasi-stable population of QPs exists at the superconducting band edge~\cite{Kaplan_qp}. The following model begins after this downconversion step. A schematic diagram of the full process can be seen in Fig.~\ref{fig:sensor_diagram}.

\subsection{QP Diffusion in Island}

Similar to what is done in~\cite{Fink_thesis,kurinsky_thesis}, we model the QP transport in the island using a 2D diffusion model. We start with the diffusion equation
\begin{align}
        \frac{\partial}{\partial t} n(\mathbf{x},t) = D_\text{island}\nabla^2n(\mathbf{x},t) - \frac{n(\mathbf{x},t)}{\tau_\text{island}} + s(\mathbf{x},t) \,,
        \label{eq:diffusion}
    \end{align}
where $n(\mathbf{x},t)$ is the QP number density, $D_\text{island}$ and $\tau_\text{island}$ are the diffusion coefficient and finite QP lifetime in the island (Al in this case), and $s(\mathbf{x},t)$ is a source of particles. 

Following the derivation from Moffatt~\cite{moffatt_thesis}, we assume that our source term $s(\mathbf{x},t)$ is a finite impulse, and we are interested in the total number of QPs absorbed. By integrating away the time degree of freedom, Eq.~\ref{eq:diffusion} becomes 
\begin{equation}
    D_\text{island}\nabla^2\phi(\mathbf{x}) - \frac{\phi(\mathbf{x})}{\tau_\text{island}} = -\sigma(\mathbf{x}) 
    \label{eq:diffusion_simp}
\end{equation}
where $\phi(\mathbf{x})$ is the scalar field defined as the time integrated particle number density and $\sigma(\mathbf{x})$ is the source density. 

\begin{figure}
    \centering
    \includegraphics[width=.9\linewidth]{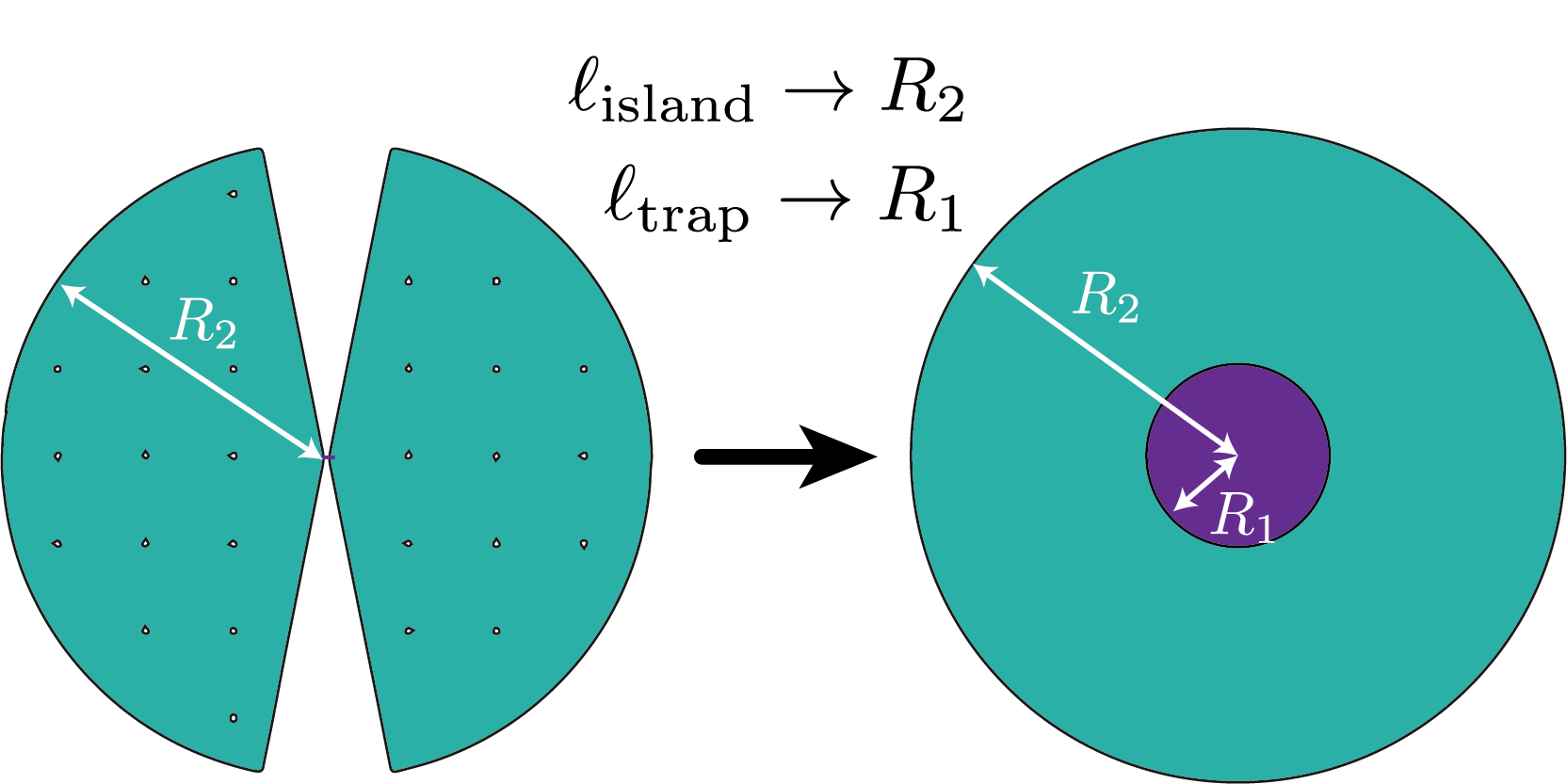}
    \caption{Schematic showing how the SQUAT sensor design (\textbf{left}) is mapped to  two concentric circles (\textbf{right)}to model the QP diffusion and trapping process. Note that the scale of $R_1$ is enlarged in the figure simply for illustrative purposes.}
    \label{fig:QP_collect_diagram}
\end{figure}

We approximate the SQUAT as two concentric circles with outer radius $R_2 = \ell_\text{island}$ and inner radius scales roughly as $R_1 = \ell_\text{trap}$ as shown in Fig.~\ref{fig:QP_collect_diagram}. Since the film thickness will be $\mathcal{O}(10-100)\,\mathrm{nm}$, and the lengths of $\ell_\text{trap}$ and $\ell_\text{island}$ with be $\mathcal{O}(10-100)\,\mu\mathrm{m}$, we can treat the QP diffusion in the islands in two dimensions. Assuming that our QP source density is homogeneous in the island (a reasonable assumption for Al where the downconversion process is orders of magnitude shorter than the QP lifetime), we can write
\begin{equation}
    \sigma = \frac{N_{\text{stable},i}}{(R_2-R_1)\pi},
\end{equation}
where $N_{\text{stable},i}$ is the total number of stable QPs after the initial downconversion process in the island. In cylindrical coordinates, Eq.~\ref{eq:diffusion_simp} becomes
\begin{equation}
    r^2\frac{\partial^2}{\partial r^2}\phi+r\frac{\partial}{\partial r}\phi - \frac{1}{L_{d,i}^2}\phi = -\frac{\sigma(r)}{D_\text{island}}
    \label{eq:diff_cyl}
\end{equation}
where $L_{d,i}\equiv \sqrt{D_\text{island}\tau_\text{island}}$ and the azimuthal angle has been suppressed due to cylindrical symmetry. 

In our model, we assert that the outer edge of $R_2$ is a perfect reflector, and that we have some boundary that scales with $R_1$ that acts as a perfect absorber. Note that this perfect absorbing boundary does not correspond to the physical boundary of the trap, but rather we define the boundary with an effective characteristic absorption length scale (defined below) to simplify the math. From these assumptions we have the boundary conditions
\begin{align}
    &\frac{\partial}{\partial r} \phi (R_2) = 0 \\
    &\frac{\partial}{\partial r} \phi(R_1) L_{a,i} = \phi(R_1),
\end{align}
where $L_{a,i}$ is defined as the absorption length, a weighting term to account for the non-perfect absorption at the trapping boundary. The absorption length can be thought of as the average distance a QP would have to travel in the trap in order to become `trapped' to make the absorption equivalent to a step function boundary with an absorption probability less than unity. By weighting $R_1$ in this way, we can treat the boundary as a step function.

The differential equation in Eq.~\ref{eq:diff_cyl} can be solved along with the boundary conditions to be expressed in terms of modified Bessel functions\footnote{In an effort to save space, only the final solution will be shown.}. The total amount of QPs that are absorbed at the boundary defined by $R_1$ will be given by
\begin{equation}
    N_{\text{absorbed},i} = 2\pi R_1 D_\text{island}\frac{\partial}{\partial r}\phi(R_1).
\end{equation}

Finally we define the fraction of QPs that get trapped in the overlapping region vs the total number of stable QPs generated after the initial downconversion process in the island as 
\begin{widetext}
\begin{equation}
    F_c^{2D} =\frac{N_{\text{absorbed},i}}{{N_{\text{stable},i}}} =  \frac{2\rho_1}{\rho_2^2-\rho_1^2}\frac{I_1(\rho_2)K_1(\rho_1)-I_1(\rho_1)K_1(\rho_2)}{I_1(\rho_2)\left[K_0(\rho_1)+\lambda_a K_1(\rho_1)\right] + K_1(\rho_2)\left[I_0(\rho_1)-\lambda_a I_1(\rho_1)\right]},
    \label{eq4:2d_diff}
\end{equation}
\end{widetext}
where $I_i$ and $K_i$ are modified Bessel functions of the first and second kind. We note that while this derivation assumed the SQUAT was a cylinder, by the assumption of cylindrical symmetry and a homogeneous source term, this result is also valid for any wedge of the cylinder as well, e.g. each island of the SQUAT.

We now address the physicality of the assumptions of perfectly reflecting and absorbing boundaries by defining the diffusion and absorption length in terms of our physical design parameters. For a sufficiently pure Al film, the diffusion is limited by the film thickness and scales linearly with the thickness $h_\text{island}$
\begin{equation}
    L_{d,i} \equiv \sqrt{D_\text{island}\tau_\text{island}} \approx \alpha_\text{island}h_\text{island}
\end{equation}
where $\alpha_\text{island}$ is a film material-specific scalar and is approximately 567 for aluminum~\cite{yen_qp_2016}. Similarly, the absorption length depends on the thickness of the island on top of the trap, the length of the trap, and the probability of absorption. As shown in~\cite{Fink_thesis,kurinsky_thesis}, this is given by

\begin{align}
    L_{a,i} \equiv \frac{D_\text{island}}{\nu_{abs}} \approx \frac{1}{p_{abs}}\frac{h_\text{island}^2}{\ell_\text{trap}},
\end{align}
 with $\ell_\text{trap}$ being the length of the overlapping region between the two materials, $\nu_{abs}$ the effective QP absorption velocity, and $p_{abs}$ the per QP transmission probability at the interface. 

We must estimate $p_{abs}$, which is a function of the island material, the trapping material, and the thickness of the trap. For this model, we propose to use AlMn for the trap and junction, because its $T_c$ can be tuned with the amount of Mn added and AlO$_x$ layers can easily be formed on it to create junctions~\cite{AlMn_NIS}. While we note that the arguments hold for other low gap materials, we use AlMn as a starting point. 

For the simplest model of QP trapping, we model the change in energy between the Al and AlMn trap as a step function between $\Delta_\text{Al}$ and $\Delta_\text{AlMn}$. In reality, this step function will be slightly smeared by the proximity effect. We assume that the QPs that enter the trap are all at energy $\Delta_\text{Al}$ (since the QP downconversion timescale is orders of magnitude faster than the average time for a QP to reach the trap). The question then becomes: what is the probability that the QP undergoes a scattering process and loses enough energy such that it cannot travel back into the Al island? We are targeting a $T_c$ for the AlMn of roughly 10 times less than that of the Al islands. The QP scattering lifetime was measured in AlMn by~\cite{AlMn_nist} to be $\sim\,65\,\mathrm{ns}$. Additionally, in~\cite{PhysRevB.79.020509} it was shown that increasing the disorder of Al films by adding Mn decreased the Al QP lifetime. Thus depending on the percent of Mn added, this scattering time could be even smaller. 

Regarding the validity of the step-function model of the Al/AlMn interface, there are two factors that we must consider. The overlapping regions between the island and the trap will have a global $T_c$ shift from the proximity effect in the Z direction, and the boundary of the overlapping region in the XY plane will have a local $T_c$ gradient from the longitudinal proximity effect~\cite{PhysRevLett.104.047003}. While the proximity effect in AlMn has not been well studied, at least for Al in the dirty limit the coherence length has been measured~\cite{Zhao_2017} to be on the order of $\xi_\text{dirty}\sim 100\,\mathrm{nm}$, which is a reasonable order of magnitude estimate for AlMn. Since our films for both the island and trap will be $\mathcal{O}(100\,\mathrm{nm})$ thick, we expect this region to be fully proximitized, and therefore we do not need to consider any spacial variation in the Z direction. Further, given the QP scattering time for AlMn, this is equivalent to an average QP scattering length of $\lambda_\text{qp} \sim \mathcal{O}(100\,\mu\mathrm{m})$. Thus for our initial designs with $\ell_\text{trap}\sim\mathcal{O}(1\,\mu\mathrm{m})$, we have the situation where
\begin{equation}
    \xi_\text{dirty}\ll \ell_\text{trap} \ll \lambda_\text{qp}.
\end{equation}
For both of these reasons, the trapping interface can be well approximated as a step function.

The scattering rate can be used to estimate the trapping probability for a given trap thickness from Eq.~\ref{eq:abs}, repeated below for completeness,
\begin{equation}
    p_{abs} = 1-\exp\left[\frac{2 t}{c_s \tau_s}\right],
\end{equation}
where $c_s$ is the sound speed in the trap, $t$ is the thickness and $\tau_s$ is the QP-phonon scattering rate. Using a trap thickness of $t=100\,\mathrm{nm}$ and an inverse scattering rate of $\tau_s=65\,\mathrm{ns}$ from~\cite{AlMn_nist}, the trapping probability for our Al/AlMn devices should be $p_{abs}\approx 6\times10^{-4}$. It is interesting to note that this value is very close to that measured by~\cite{yen_thesis, yen_qp_2016}, who found that the trapping probability for Al/W traps is $p_{abs}\approx 10^{-4}$. For this model we will thus take the conservative approach and use $p_{abs}\approx 10^{-4}$ in our calculations.

\subsection{Quasiparticle Multiplication}

When the QPs enter the trap they will have an energy of roughly $\Delta_\text{island}$ (see Fig.~\ref{fig:sensor_diagram} b). Once in the lower $T_c$ trap however, these QPs will undergo a downconversion step into a lower energy QP population and athermal phonons. During the downconversion process, a fraction of the athermal phonons created have less energy than $2\Delta_\text{trap}$ and will be lost into the substrate. The energy that remains in the QP system ($\eta_\text{QP}$) depends on the incident energy of the original particle, in this case of energy $\Delta_\text{island}$. If the initial QP has energy of $2\Delta_\text{trap}$, then $\eta_\text{QP}$ is approximately unity. This is because the QPs are already roughly at the band edge and will not undergo downconversion. If the incident particle has energy greater than $2\Delta_\text{trap}$ or less than $4\Delta_\text{trap}$, $\eta_\text{QP}$ decreases significantly, because the only kinimatically allowed phonons released will be sub-gap, and thus not contribute to the QP population, roughly half the energy will go into sub-gap phonons. For incident energies above this range, $\eta_\text{QP}$ tends to asymptote to a fixed value, this value depends in part on QP lifetimes and above-gap phonon loss mechanisms that are material dependent. A simple simulation of this is shown in Fig.~\ref{fig:QP_gain_eff}. To simulate this, QP of energy $\Delta_\text{island}/\Delta_\text{trap}$ enters the trap and undergoes spontaneous downconversion, where a random percentage of its energy is given to a phonon and the rest remains with the QP, the above rules are then used to determine if the phonon is energetic enough break a cooper-pair into two new QPs. This process is repeated until all QPs are no longer energetic enough to generate pair-breaking phonons. In this simulation, it is assumed that both the QP and phonon lifetime is large compared with the scattering times.

While some portion of the energy is lost to sub-gap phonons, there is ultimately a QP number enhancement that depends on the ratio of the SC gaps. The QP number gain is given roughly by~\cite{booth_qp_trap} 
\begin{align}
    G_{QP} \approx \eta_\text{QP}\frac{\Delta_\text{island}}{\Delta_\text{trap}}\,,
    \label{eq:qp_gain}
\end{align}
which, as shown in Fig.~\ref{fig:QP_gain_eff}, when the ratio becomes greater than $\sim 3$, $\eta_\text{QP}$ starts to become constant at a value of roughly $60\%$, and the gain scales linearly with the ratio. For our targeted island-to-trap gap ratio of $\sim 10$, we should expect a QP number gain of about 7. We note that while this calculation assumes that all of the downconverted phonons that are energetic enough to generate new QPs will do so, we have the freedom in our design to dramatically increase the trapping thickness to minimize additional phonon losses and maximize the number of generated QPs. This value should be thought of as the upper bound for the maximum achievable gain.

\subsection{Quasiparticle Diffusion in Trap}

\begin{figure}
    \centering
\includegraphics[width=.9\linewidth]{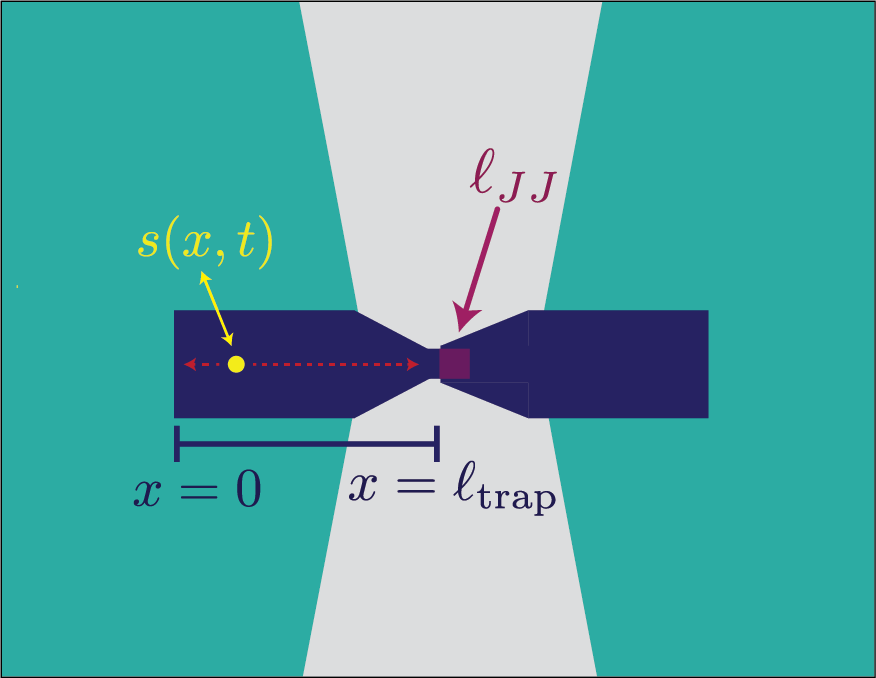}
    \caption{Schematic diagram of the trap and junction region (not to scale), showing the junction area, the trap, and the source term of QPs, $s(x,t)$.}
    \label{fig:QP_drift_in_trap}
\end{figure}

At this point, a stable population of QPs has developed in the trap. We then model the trap QP transport using a 1D diffusion model, which is justified in this case since the length of this region is much larger than the width or film thickness. Since the QPs now have an energy much less than $2\Delta_\text{island}$, we treat the boundary between the edge of the trap and the island ($x=0$ in Fig.~\ref{fig:QP_drift_in_trap}) as perfectly reflecting. We treat the junction as a perfect absorber (where we consider a `tunneled' QP to be `absorbed'), and wrap the QP tunneling probability into the definition of the absorption length, similar to the two dimensional case above for the islands. 

In this case, we treat our source term in the diffusion equation as a delta function, both spatially and temporally, and Eq.~\ref{eq:diffusion} can be expressed in one dimension as 
\begin{equation}
    \left(\partial_x^2 - \frac{1}{L_{d,t}}^2\right)\phi = -\frac{N_\text{stable, t}}{D_\text{trap}}\delta(x-\ell_\text{trap}),
\end{equation}

Where again, $N_{\text{stable},t}$ is the total number of stable QPs after the downconversion/multiplication process in the trap. The above equation is solvable under the same boundary conditions as for the two dimensional case above, resulting in a fraction of tunneled QPs to the total number of stable QPs generated in the trap of
\begin{equation}
    F_c^{1D} = \frac{\lambda_d}{\lambda_a/\lambda_d + \coth\left(\frac{1}{\lambda_d}\right)},
    \label{eq:1d_diff_raw}
\end{equation}
where $\lambda_d = L_{d,t}/\ell_\text{trap}$ and $\lambda_a = L_{a,t}/\ell_\text{trap}$.

Similar to before, we now account for the non-ideal absorption by relating the absorption and diffusion lengths to physical parameters. The diffusion in the trap will again be limited by surface effects and thus
\begin{align}
    L_{d,t} \approx \alpha_\text{trap}h_\text{trap},
\end{align}
where $h_\text{trap}$ is the trap thickness and $\alpha_\text{trap}$ is the scalar quantifying the diffusion in the thickness limited limit. We can describe the absorption the same way as for the two dimensional case, replacing absorption with tunneling effects,
\begin{equation}
    \lambda_a = \frac{1}{p_\text{tunnel}}\frac{h_\text{trap}^2}{\ell_{JJ}\ell_\text{trap}}
\end{equation}
where $\ell_\text{JJ}$ is the length of the Josephson junction and $p_\text{tunnel}$ is the per-QP tunneling probability at the junction. Combining these definitions, we can write Eq.~\ref{eq:1d_diff_raw} as
\begin{equation}
    F_c^{1D} \approx \left[1+\frac{\ell_\text{trap}}{\ell_{JJ}}\frac{1}{\alpha_\text{trap}^2p_\text{tunnel}}\right]^{-1} \,,
    \label{eq:1d_diff}
\end{equation}
where we have expanded the hyperbolic cotangent term, because $1/\lambda_d$ can be made much less than 1 with reasonable design parameters. 

For illustrative purposes of our model, we assume AlMn for our trapping material. Since the relative amount of Mn in the AlMn should be very low, the QP scattering diffusion length should be close to that of a typical Al thin film. As such, we use a value of $\alpha_\text{trap}\approx 570$ as measured in~\cite{yen_qp_2016}. From~\cite{QP_rates} we estimate the QP tunneling probability to be $p_{tunnel}\approx 1-10 \,(\times10^{-6})$ for pure Al. For the AlMn in this model, we adopt the conservative lower bound of $1\times10^{-6}$. 

This step is where the true utility of this design comes into fruition. While the QP tunneling probability across the Josephson junction is very small---requiring a QP to impinge upon the junction $10^6$ times on average before it tunnels---this setback is counteracted by confining the QPs to a volume that is at least 5 orders of magnitude smaller than the islands.

\subsection{Signal Gain from Multiple Tunneling Events}

Since there is no potential difference across the junction, QPs near the junction will have multiple chances to tunnel before recombining. Using simple geometric arguments, we estimate the characteristic time it takes a QP in the trap to reach the junction as
\begin{align}
    \tau_{qp\to JJ} \approx \frac{4 \text{V}_\text{trap}}{\ell_\text{JJ}^2v_{QP}}
    \label{eq:tau_tun}
\end{align}
where $\text{V}_\text{trap}$ is the volume of the trap and $\ell_\text{JJ}$ is the length of a single side of the square Josephson junction, and $v_{QP}$ is the QP velocity at the operating temperature. We can then define the average number of tunnels per QP as
\begin{equation}
    \left<n_\text{tunnel}\right> \approx p_\text{tunnel}\frac{\tau_\text{life}}{\tau_{qp\to JJ}}
    \label{eq:n_qp}
\end{equation}
where $\tau_\text{life}$ is the QP lifetime in the junction material.  While this term depends on many material parameters, it is reasonable to expect each QP will be measured $\mathcal{O}(10-100)$ times for an $\mathcal{O}(1-10\,\mu\mathrm{m})$ sized trapping region. However, since this effect needs to be studied, for this current model we make the conservative assumption that $\left<n_\text{tunnel}\right>=1$.

\subsection{Total Quasiparticle Collection Efficiency}

The final fraction of `collected QPs' (where a collected QP is defined as a tunneling event) starting with a stable QP population in the island is thus given by the product of Eqs.~[\ref{eq4:2d_diff}]\,[\ref{eq:qp_gain}]\,[\ref{eq:1d_diff}]\,[\ref{eq:n_qp}]
\begin{equation}
    F_\text{collect} = F^{2D}_\text{island}\times G_{QP} \times F^{1D}_\text{trap} \times \left<n_\text{tunnel}\right>.
    \label{eq:total_QP_collect}
\end{equation}

\begin{figure*}
\centering
\includegraphics[width=\linewidth]{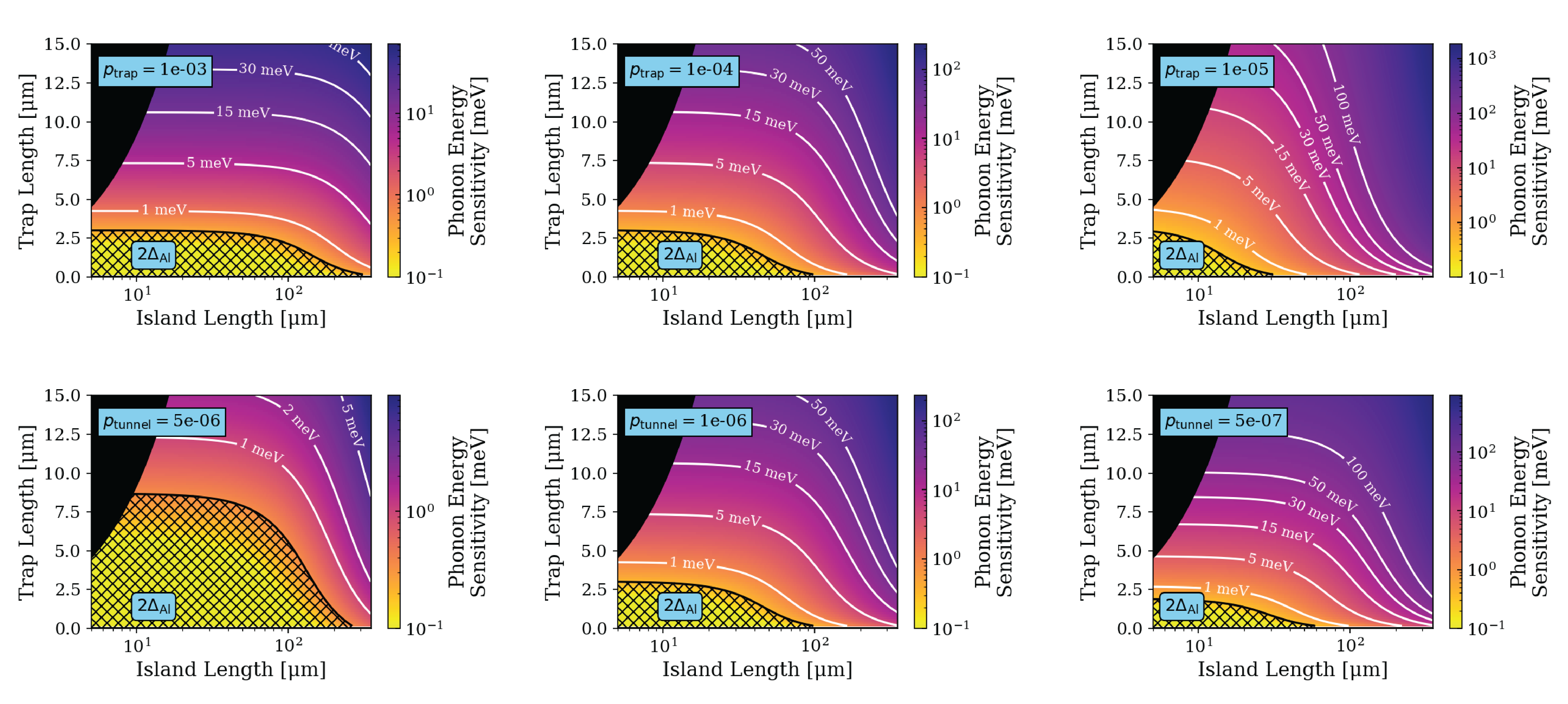}
\caption{Simulated phonon energy resolution for single QP sensitivity as a function of island and trap length assuming negligible change in fill factor as a function of island length. Shown in the hatched region is twice the superconducting gap of Al, and the blacked out regions represent non-physical design parameters. \textbf{Top}, Resolution as a function of $p_\text{trap}$ with fixed $p_\text{tunnel}=1\times 10^{-6}$ for $p_\text{trap} = [1\times 10^{-3}, 1 \times 10^{-4}, 1 \times 10^{-5}]$ from left to right. \textbf{Bottom}, Resolution as a function of $p_\text{tunnel}$ with fixed $p_\text{trap}=1\times 10^{-4}$ for $p_\text{tunnel} = [5\times 10^{-6}, 1 \times 10^{-6}, 5 \times 10^{-7}]$ from left to right.}
\label{fig:energy_res_vary_params}
\end{figure*}

An example plot of how this collection fraction depends on the model parameters is shown in Fig.~\ref{fig:QP_gain_eff} for the parameters given in the main text. Fig.~\ref{fig:energy_res_vary_params} shows how the final energy resolution depends on the trapping and tunneling probabilities.

\section{Sensor Bandwidth}
\label{appendix:BW}

While in previous sections we have focused on only the low energy threshold of the SQUAT, we expect the dynamic range of the sensor to be limited primarily by the readout bandwidth. This can be understood intuitively as follows.  The number of QPs produced in the SQUAT will scale linearly with the total energy $E$ of the original photon/phonon. When a large amount of QPs are near the junction, the average time between individual QP tunneling events will scale with the total number of QPs. Consequently, this means that the tunneling rate is a function of particle energy. In a practical readout mode, the SQUAT will have a finite rate a which it can make a measurement. This finite readout bandwidth implies that sequential tunneling events that take place faster than the readout rate of $1/2\tau_\text{BW}$ will likely not be measured. As such, the total number of QPs measured will no longer be linear with the original energy above a certain threshold, as set by the readout bandwidth. In this appendix we now derive where this threshold is, and how it sets the dynamic range of the SQUAT.

The SQUAT is sensitive to single QP tunneling events, and thus the metric we are concerned with calculating is the typical time between tunneling events as a function of incident particle energy. We consider first the case of a single photon or phonon creating a population of QPs near the junction of the SQUAT. While this is technically a two stage process---the QPs must travel through the island before getting trapped and then must tunnel across the junction---the trapping time is at least four orders of magnitude faster than the tunneling time. As such, we can ignore the trapping time scale and consider only a single time scale of $\tau_\text{tunnel}$.

Since the number of QPs generated from a single photon or phonon interaction is governed by Poissonian statistics, this means that the expected time between QP tunneling events will be well described by an exponential distribution. Noting that there is a one-to-one correspondence between `time between tunneling events' and `measured QPs'; we can estimate the percentage of measured QPs to be the same as the percentage of tunnel-to-tunnel times greater than $2\tau_\text{BW}$. Thus the measurement efficiency due to finite sensor bandwidth as a function of the event energy is
\begin{equation}
    \varepsilon_\text{BW}(E) = \int_{2\tau_{BW}}^{\infty} f_\text{exponential}(t, E)\,dt,
    \label{eq:BW_eff}
\end{equation}
where $f_\text{exponential}(t, E)$ is the exponential distribution given by
\begin{equation}
    f_\text{exponential}(t, E) = \frac{1}{\tau_\text{tunnel}(E)}e^{\left(-\frac{t}{\tau_\text{tunnel}(E)}\right)}.
    \label{eq:singe_pole_fraction}
\end{equation}

To estimate the efficiency given by Eq.~\ref{eq:BW_eff}, we first calculate $\tau_\text{tunnel}$. Considering a single QP in the trap, the characteristic time that it will take to diffuse to the trap is given by Eq.~\ref{eq:tau_tun}. The average time between tunneling events ($\tau_\text{tunnel}$) will scale with both the total number of QPs ($N_\text{QP}$) and the average number of tunnels per QP ($\left<n_\text{tunnel}\right>$), as

\begin{align}
    \tau_\text{tunnel} =& \frac{\tau_{qp\to JJ}}{N_\text{QP}(E)\left<n_\text{tunnel}\right>}\\
    =& \frac{\tau_{qp\to JJ}^2}{p_\text{tunnel} N_\text{QP}(E) \tau_\text{life}},
\end{align}
where the definition of $\left<n_\text{tunnel}\right>$ from Eq.~\ref{eq:n_qp} was used in the second line. 

While in general there will be sequential tunneling events that occur faster than the inverse readout bandwidth, all of these events are not completely lost. In the worst case scenario where a large fraction of sequential events tunnel faster than the readout rate, on average, we will simply measure a tunneling rate limited to $1/2\tau_\text{BW}$. Therefore our readout efficiency in Eq.~\ref{eq:BW_eff} will be

\begin{equation}
    \varepsilon_\text{BW}(E) =\begin{cases} \exp{\left(-\frac{2\tau_\text{BW}}{\tau(E)_\text{tunnel}}\right)}, \quad &\tau_\text{tunnel} > 2\tau_\text{BW}\\

    \exp{\left(-1\right)}, \quad &\tau_\text{tunnel} \leq 2\tau_\text{BW}
    \end{cases}
    \label{eq:BW_eff_eval}
\end{equation}
 where $\varepsilon_\text{BW}$ is defined as the number of QP tunneling events measured, divided by the total number of QPs created in the trap.  While the efficiency at larger energies will be reduced, we will not lose sensitivity to these events entirely. The expression in Eq.~\ref{eq:BW_eff_eval} represents the case of single photon or phonon absorption by the SQUAT, which can be thought of as a delta function impulse being convolved with an exponential distribution. However in general, for phonon absorption there will typically be a Poissonian phonon population in the substrate as well. In this case, Eq.~\ref{eq:BW_eff} must now integrate the convolution of two exponential distributions, one with an average time constant of $\tau_\text{tunnel}$ and one with an average of $\tau_\text{phonon}$ given in Eq.~\ref{eq:tau_ph}. In this case, the efficiency due to the bandwidth becomes

\begin{equation}
    \varepsilon_\text{BW}(E)^\text{phonon} = A\left(\frac{e^{-2\lambda_2\tau_\text{BW}}}{\lambda_2}-\frac{e^{-2\lambda_1\tau_\text{BW}}}{\lambda_1}\right),
    \label{eq:phonon_BW}
\end{equation}
where 
\begin{align}
    A=& \frac{\lambda_1 \lambda_2}{\lambda_1-\lambda_2},\\
    \lambda_1 =& \frac{1}{\tau_\text{phonon}},\\
    \lambda_2 =& \frac{1}{\tau_\text{tunnel}}, \quad \tau_\text{tunnel} > 2\tau_\text{BW}\\
    \lambda_2 =& \frac{1}{\tau_\text{BW}}, \quad \tau_\text{tunnel} \leq 2\tau_\text{BW}
\end{align}

\begin{figure}
    \centering
\includegraphics[width=.9\linewidth]{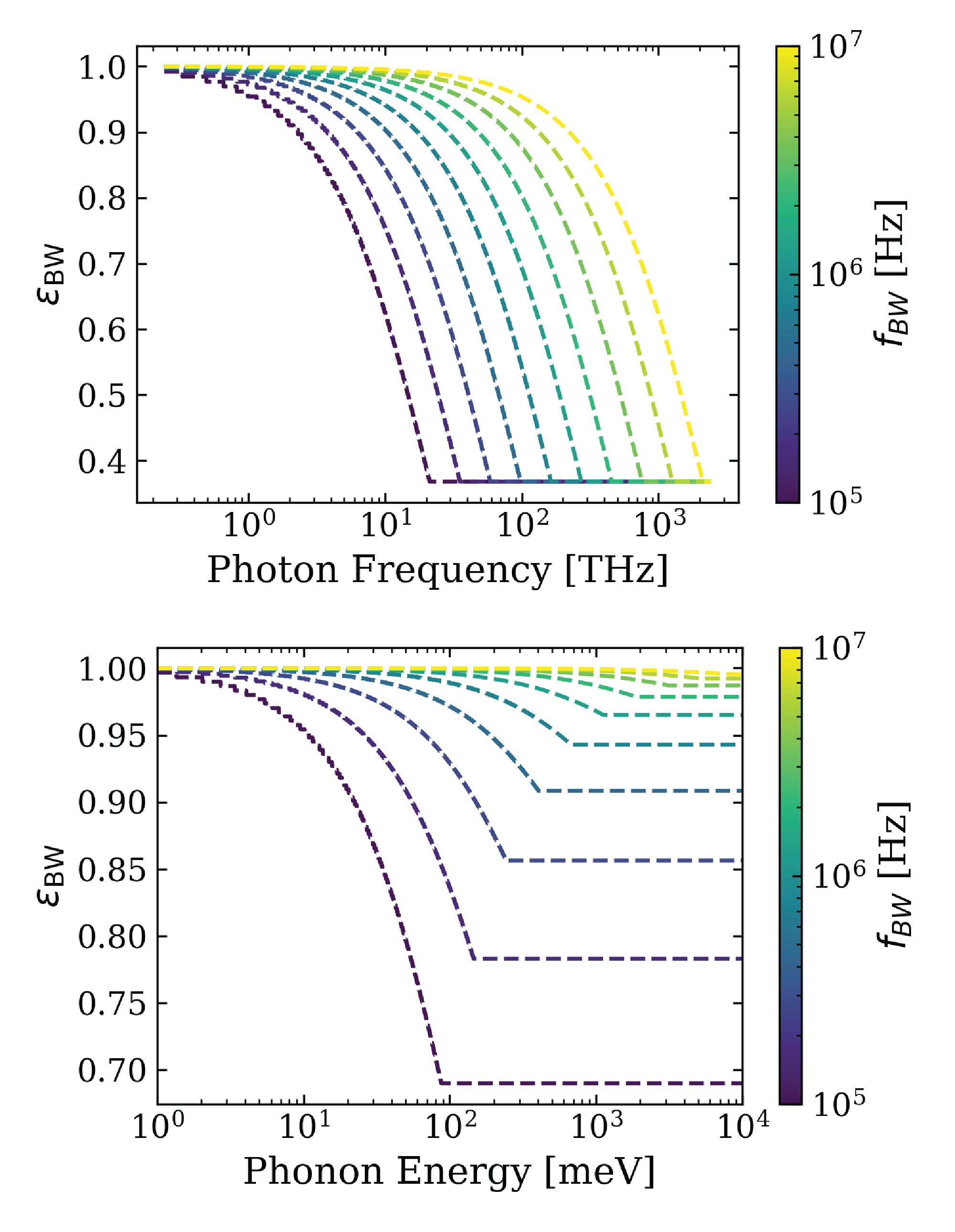}
    \caption{\textbf{Top:} QP number efficiency, $\varepsilon_\text{BW}$, for the case of a direct energy deposit into the SQUAT, e.g. singe photon absorption, from Eq.~\ref{eq:BW_eff_eval}. \textbf{Bottom:} Estimated QP number efficiency for phonon absorption due to the finite readout bandwidth for a geometry of $\ell_\text{island} = 100\,\mu\mathrm{m}$ and $\ell_\text{trap} = 4\,\mu\mathrm{m}$, assuming $\tau_\text{phonon}=2\,\mu\mathrm{s}$. $\varepsilon_\text{BW}$ is defined as the number of QP tunneling events measured, divided by the total number of QPs created in the trap.}
    \label{fig:BW_eff}
\end{figure}

\begin{figure}
    \centering
\includegraphics[width=1\linewidth]{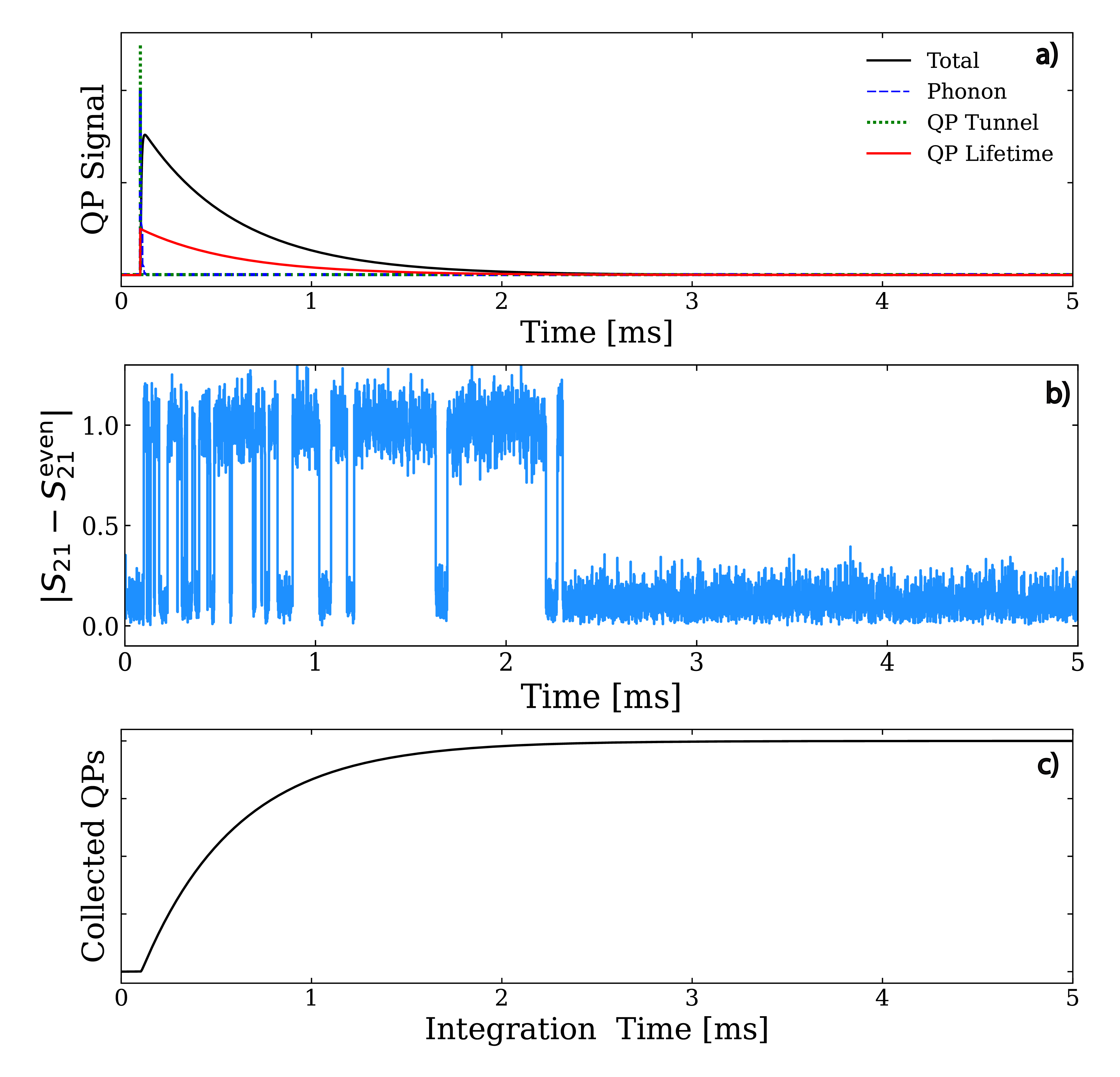}
    \caption{Example of characteristic event signal. \textbf{a)} Analytic components of a typical phonon event. Note the component pulses are arbitrarily scaled. \textbf{b)} Simulated parity switching for the same pulse shown above. \textbf{c)} Total collected QPs as a function of integration time (arbitrary units).}
    \label{fig:phonon_signals}
\end{figure}

A plot of Eq.~\ref{eq:phonon_BW} is shown in Fig.~\ref{fig:BW_eff}, which gives an idea of the linearity, or dynamic range of the SQUAT as a function of phonon energy for a variety of different readout bandwidths. From Fig.~\ref{fig:BW_eff} we can see that in the case that the phonon pulse is slower than the tunneling time and inverse readout bandwidth, the SQUAT readout efficiency will saturate. A simulated typical phonon pulse is shown in Fig.~\ref{fig:phonon_signals} as well as a depiction of what the measurement will physically look like in terms of the tunneling rate observed by the sensor. 


\section{SQUAT geometric tuning and simulation}

To achieve the optimal device parameters for a given readout method (as detailed in Appendix~\ref{appendix:fidelity}), we must tune the geometry of the qubit and its coupling to the feedline.  In this appendix we show how tuning various geometric parameters affects the readout parameters as simulated with ANSYS HFSS, a finite element RF modeling software \cite{ansys}.

\begin{figure}
    \centering
\includegraphics[width=.9\linewidth]{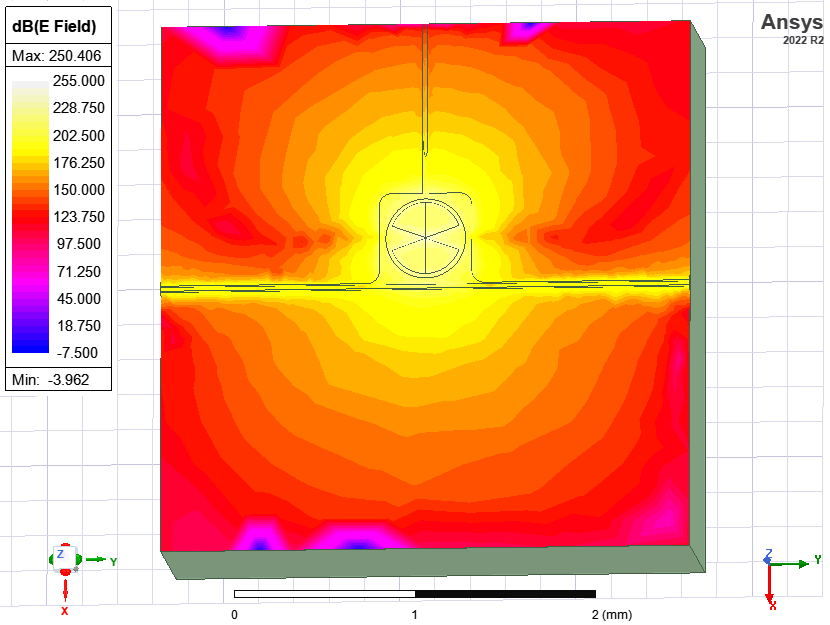}
    \caption{Electric fields of a SQUAT that is driven at the qubit resonance as simulated in ANSYS HFSS.}
    \label{fig:squatField}
\end{figure}

Each qubit is measured with a modal network simulation type.  Just as the SQUATs will be in actual measurements, they are stimulated with a transmission measurement along the feedline.  This is achieved in simulation with ports that inject modes at swept frequencies across the qubit resonance.  An example electric field plot for a qubit driven on resonance is shown in Fig.~\ref{fig:squatField}.

\begin{figure*}
    \centering
\includegraphics[width=0.45\linewidth]{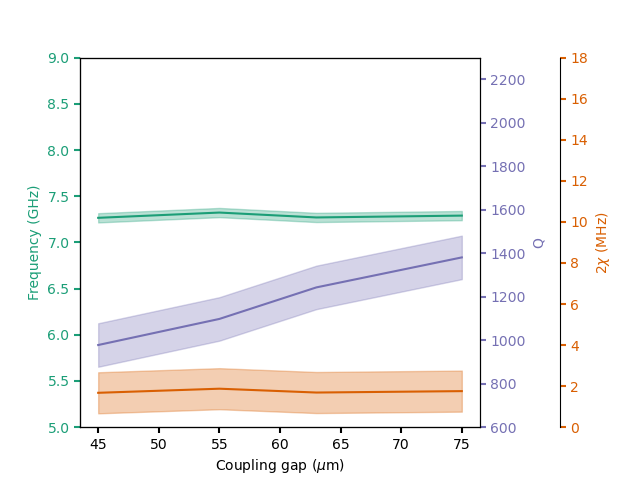}
\includegraphics[width=0.45\linewidth]{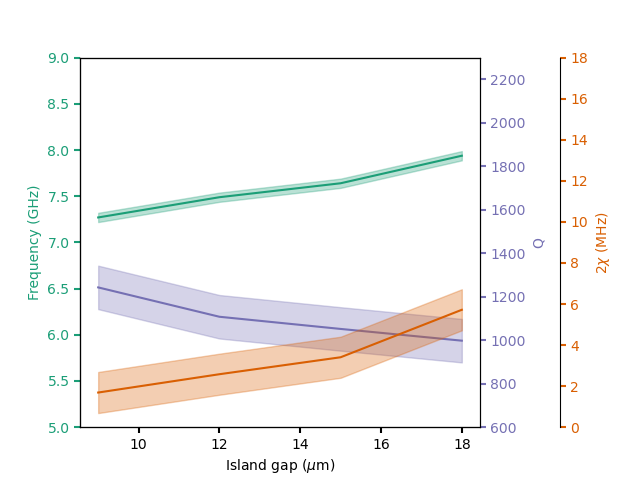}\\
\includegraphics[width=0.45\linewidth]{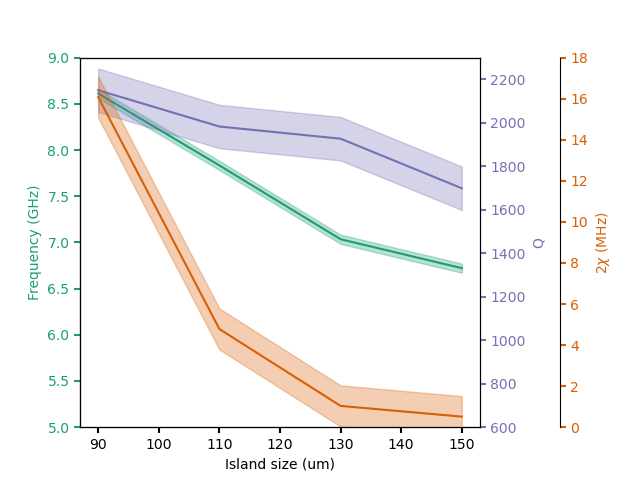}
\includegraphics[width=0.45\linewidth]{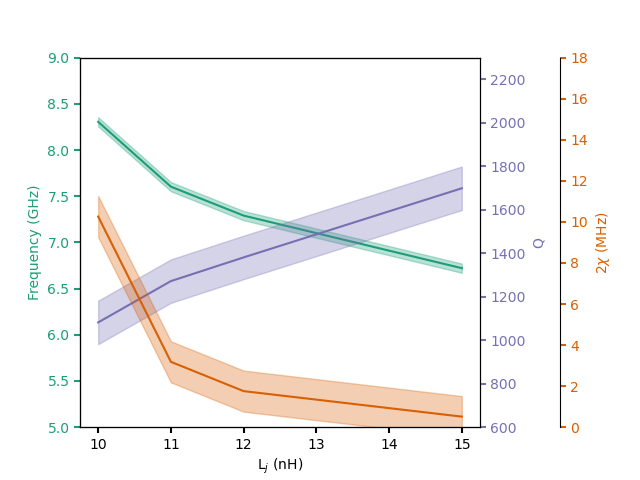}
    \caption{SQUAT parameters vary with the device geometry. In each plot, only one geometric parameter was varied. The uncertainty bands reflect the level of variation within one simulation method.  See the text for further discussion of systematic uncertainty. \textbf{Top left}, variation of the coupling distance between the qubit and the feedline and ground plane.  \textbf{Top right}, variation of the gap between the qubit islands. \textbf{Bottom left}, variation of the size of the qubit islands. \textbf{Bottom right}, variation of the Josephson inductance, which is a function of the junction size and thickness. }
    \label{fig:geomTrends}
\end{figure*}

The trends in frequency, quality factor, and charge dispersion resulting from the HFSS simulations are shown in Fig.~\ref{fig:geomTrends}.  The uncertainties shown there reflect how much variation in those parameters we expect based on a single simulation method, and do not include systematic uncertainties associated with differences between simulation and the real device.  We expect the frequencies and quality factors to be fairly accurate, but the dispersion's exponential dependence on $E_J/E_C$ (see Eq.~\ref{eq:dispersion}) means that small discrepancies in the simulated versus actual $E_J$ and $E_C$ can lead to large differences in $2\chi$.

To see the magnitude of this effect, we ran an ANSYS Maxwell simulation to get another estimate for $E_C$ for some of the data points shown in the plots.  The difference in $E_C$ was up to a factor of $\sim2$, leading to shifts of $2\chi$ by up to two orders of magnitude.  However, although the two simulation methods give different results, the overall dispersion trends with geometry should be consistent between the methods.  Thus the $2\chi$ curves in Fig.~\ref{fig:geomTrends} should be interpreted for the geometric trends rather than for the values themselves.  Once devices have been fabricated and tested, we will benchmark our simulations to the measured values.

\end{document}